\documentclass[aps,prb,twocolumn,amsmath,amssymb,floatfix]{revtex4-1}
\usepackage{tabularx}
\usepackage{bm}
\usepackage{euscript}
\usepackage{graphicx}
\usepackage{color}
\usepackage{amsfonts}
\usepackage{exscale}
\usepackage{amsbsy}
\usepackage{subfigure}
\usepackage{textcomp}

\newcommand{\nn}{\nonumber}

\newcommand{\be}{\begin{eqnarray}}
\newcommand{\ee}{\end{eqnarray}}
\newcommand{\tr}{\textrm{tr}}
\newcommand{\CL}{{\cal L}}
\newcommand{\CO}{{\cal O}}

\begin{document}

\title{Non-Fermi liquid fixed point in a Wilsonian theory of quantum critical metals}
\author{A. Liam Fitzpatrick$^{\bar \psi}$, Shamit Kachru$^{{\bar \psi},\psi}$, Jared Kaplan$^{{\bar \psi},\phi}$, S. Raghu$^{{\bar \psi},\psi}$}
\affiliation{$^{\bar \psi}$Stanford Institute for Theoretical Physics, Stanford University, Stanford, California 94305, USA}
\affiliation{$^\psi$SLAC National Accelerator Laboratory, 2575 Sand Hill Road, Menlo Park, CA 94025}
\affiliation{$^\phi$Department of Physics and Astronomy, Johns Hopkins University, Baltimore, MD 21218}

\date{\today}

\begin{abstract}
We study the problem of disorder-free metals near a continuous quantum critical point.  We depart from the standard paradigm~\cite{Hertz1976, Millis1993}, and treat both fermions and bosons ({\it i.e.} order parameter fields) on equal footing.  We construct a Wilsonian effective field theory that integrates out {\it only high energy boson and fermion modes}.  Below the upper critical dimension of the theory ($d=3$ spatial dimensions),  we find  new fixed points in which the bosons are described by the Wilson-Fisher fixed point and are coupled to a non-Fermi liquid metal.  
We describe subtleties with the renormalization group flow of four-Fermi interactions, which
can be surmounted in a controlled large N limit.
In this limit, we find that the theory has no superconducting instability.

\end{abstract}

\maketitle

\section{Introduction}
Landau Fermi liquid theory\cite{Landau1957} is a remarkably successful framework that explains how a metal can remain a stable phase of matter over a wide range of energy scales, despite having infinitely many gapless excitations.  From the  modern perspective of effective field theory, a Fermi liquid is governed by a renormalization group (RG) fixed point in which most interactions are irrelevant, due to the kinematic constraints  imposed by  a Fermi surface\cite{Shankar1991,Polchinski1992, Shankar1994}.    
Indeed, the only way in which a disorder-free Fermi liquid can be destabilized is by effective attractive interactions, which lead to superconductivity\cite{Shankar1991,Polchinski1992, Shankar1994}.

A significant fraction of highly correlated materials, however, are not well described by the Fermi liquid paradigm\cite{Lohneysen2007,Stewart2001}.  A key challenge remains to construct controlled effective field theories of these ``non-Fermi liquid" metals that encapsulate their universal properties and describe their stability.  It is believed that an essential ingredient for non-Fermi liquid behavior is the presence of  additional gapless degrees of freedom (bosons tuned to criticality, or  unscreened  gauge fields are two examples) that act as a source of dissipation for the otherwise weakly interacting fermions of the metal.   Many have postulated that the resulting strongly coupled system can capture much of the phenomenology of highly correlated electron materials albeit in a vastly simplified  context\cite{Holstein, Varma, Altshuler1994, Nayak1994, Polchinski1994, Chakravarty1995, Oganesyan2001,Varma2002, FradkinExtra, Senthil2009, Lee2009,Metlitski2010,Mross2010,Lee2013}.

Our focus here will be on quantum critical metals, which  are  described by a Lagrangian containing, in addition to the fermions of the metal, bosonic order paramer fields whose mass is tuned to zero at a quantum critical point.    
The standard paradigm for understanding quantum critical metals\cite{Hertz1976, Millis1993}  involves integrating out all fermionic excitations, including those modes that lie on the Fermi surface.   In a metal, this procedure is dangerous: integrating out gapless modes on the Fermi surface will give rise to 
non-analytic, and even singular effective interactions among the bosons\cite{Abanov2004, Belitz2005}.   A more systematic treatment of such phenomena would invoke a Wilsonian coarse-graining procedure in which only high energy modes are integrated out.  
A  Wilsonian effective field theory  can never  generate singular or non-analytic corrections to the action and can in principle be analyzed in a controlled fashion.

To date, all descriptions of non-Fermi liquids involve  effective theories based on non-analytic actions of one form or another\cite{Holstein, Varma, Altshuler1994, Nayak1994, Polchinski1994, Chakravarty1995, Oganesyan2001,Varma2002, Lee2009, Metlitski2010, Mross2010};  they can only be obtained by integrating out gapless modes.  By contrast, we are motivated here by asking whether non-Fermi liquid fixed points can arise in Wilsonian effective field theories.  By explicit construction, we show that this is indeed the case, which therefore places the notion of a non-Fermi liquid fixed point on firmer ground.  In the vicinity of the upper-critical dimension,   which as we discuss below is $d=3$ spatial dimensions for the class of transitions studied here, we find  new fixed points in  which the bosons are described by a Wilson-Fisher fixed point and are coupled to a non-Fermi liquid.  

\begin{figure}
\begin{center}
\includegraphics[width=0.35\textwidth]{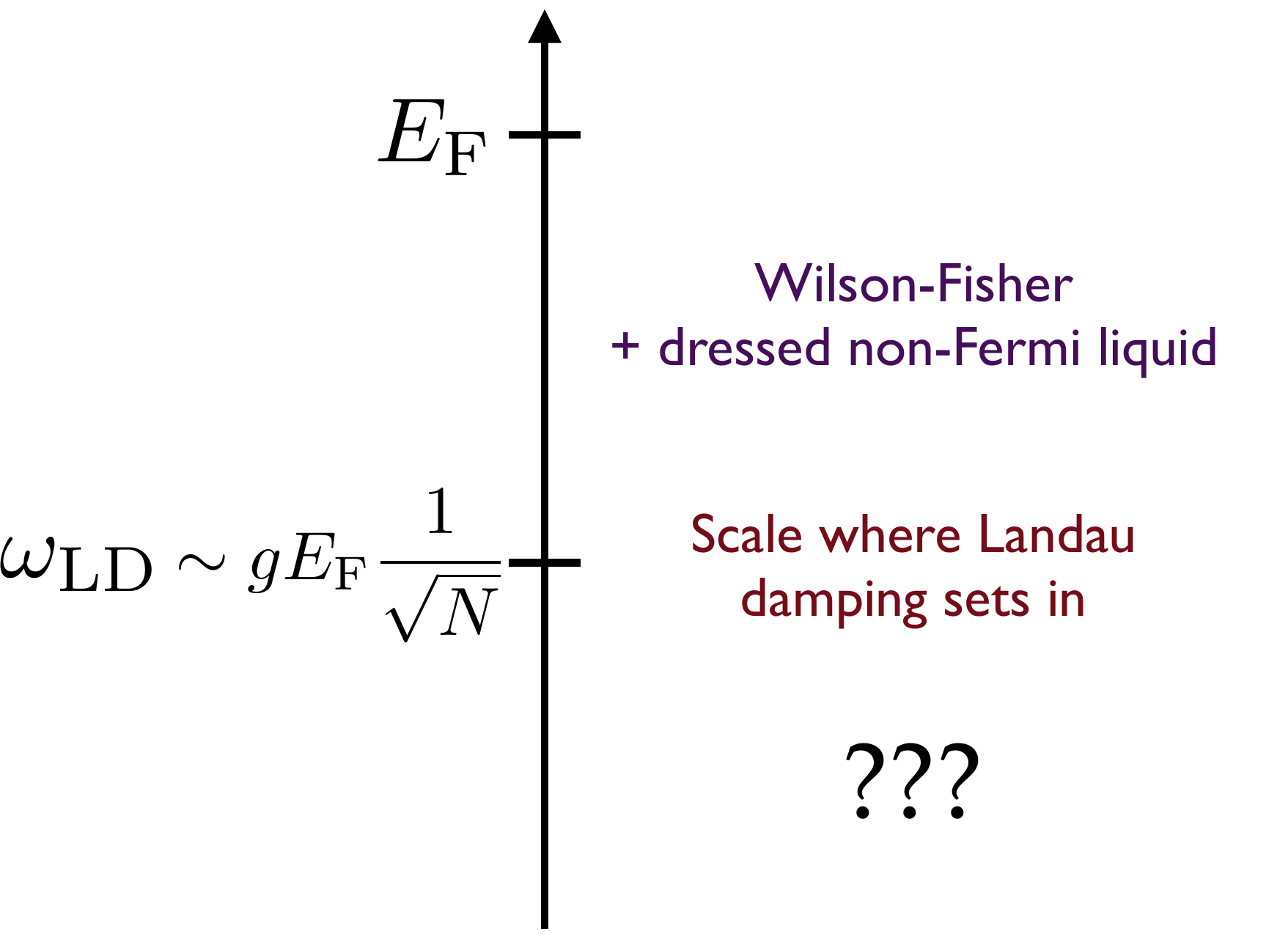}
\end{center}
\caption{ This figure depicts the regime of energy scales over which our description is controlled.  The physics below the parametrically low scale of Landau damping remains to be understood. 
 }
\label{fig:ShamitFigure}
\end{figure}

 Non-Fermi liquid fixed points of Fermi surfaces coupled to Landau damped U$(1)$ gauge bosons were first studied in an expansion about the upper-critical dimension in [\onlinecite{Chakravarty1995}].  We follow a similar approach, but start instead with a UV fixed point corresponding to a Fermi liquid coupled to undamped critical order parameter fields.  In a large $N$ limit to be discussed in detail, the scaling trajectories away from the UV fixed point lead unambiguously to the non-Fermi liquid fixed point obtained here;   the properties associated with this non-Fermi liquid fixed point are different than the predictions of the standard approach to the problem\cite{Sachdev}.  In this limit, the metal remains also stable against the presence of infinitessimal attractive interactions at the non-Fermi liquid fixed point.  For small N, however, there are IR singularities associated with Landau damping, as well as interactions in the BCS channel which may cause the scaling trajectories to flow away from the fixed point.  Thus, for small N, the fixed point described below describes  the intermediate asymptotic behavior, above energy scales that can be parametrically suppressed in the expansions to be considered here (see Fig. \ref{fig:ShamitFigure}).  

 In this paper, we will restrict our analysis to Pomeranchuk instabilities, a classic and well-studied set of phase transitions in condensed matter physics, in which rotational symmetry is broken whereas translation symmetry remains preserved.  In the case of continuous Pomeranchuk transitions, the bosons condense at zero momentum and therefore couple to fermions at every point of the Fermi surface.  There is growing experimental evidence that such transitions have been observed in  several families of highly correlated materials including the cuprate superconductors as well as in heavy fermion compounds\cite{Fradkin2010}.  A similar treatment  can be applied to the case of quantum critical phenomena associated with the density wave orders.  We will consider these transitions in a separate publication.

The paper is organized as follows.  In section 2, we construct a scaling theory that treats both low energy bosons and fermions on an equal footing, which manages to capture the correct behavior of both the fermion and boson degrees of freedom when they are decoupled from one another.  In section 3, we describe our renormalization group strategy and construct a non-Fermi liquid fixed point that governs the theory in absence of four-Fermi interactions.  We describe the correlation functions of both the boson and fermion degrees of freedom at the non-Fermi liquid fixed point; they differ from the results obtained in alternative treatments.  In \S4, we re-introduce the four-Fermi interactions and describe subtleties associated with log-squared divergences that arise in their presence.  
In \S5, we discuss controlled large $N$ theories  where the subtleties of \S4 do not arise, and we find fixed points which generalize
those of \S3 to include four-Fermi interactions.  We show that these fixed points have no superconducting instabilities.  We close with a discussion of open issues in \S6.
Explicit calculations which we refer to in the main body are presented in several appendices.

\section{Effective action and scaling analysis}
In the standard description of quantum critical points in metals\cite{Hertz1976}, one starts with a theory involving fermion fields $\psi_{\sigma}$ with spin $\sigma = \uparrow, \downarrow$ interacting at short distances with strong repulsive forces.   
These interactions are decoupled by an auxiliary boson field $\phi$ representing a fermion bilinear, and the partition function is obtained by averaging over all possible values of both the fermion and boson fields.  
Initially, the auxiliary field has no dynamics and is massive.  However, as high energy modes of the material of interest are integrated out, radiative corrections induce dynamics for the bosons.  

 In a Wilsonian theory, the dynamics are encapsulated only in {\it local, analytic corrections to the bare action}.  This mode elimination is continued until eventually, the UV cutoff $\Lambda \ll E_F$ represents the scale up to which the quasiparticle kinetic energy $\epsilon(\bm k)$ can be linearized about the Fermi level.  At these low energies, and in the vicinity of the quantum critical point where the field $\phi$ condenses, it is legitimate to view $\phi$  as an independent, emergent fluctuating field.  
The resulting effective low energy Euclidean action consists of a purely fermionic term, a purely bosonic term and a Yukawa coupling  between bosons and fermions:
\begin{eqnarray}
\label{action}
\mathcal S &=& \int d \tau \int d^d x \  \mathcal  L = S_{\psi} + S_{\phi} + S_{\psi-\phi} \nonumber \\
\mathcal L_{\psi} &=&   \bar \psi_{\sigma} \left[ \partial_{\tau}  + \mu -\epsilon(i  \nabla)  \right] \psi_{ \sigma} + \lambda_{\psi} \bar \psi_{\sigma} \bar \psi_{\sigma'} \psi_{\sigma'} \psi_{\sigma} \nonumber  \\
\mathcal L_{\phi} &=& m_{\phi}^2   \phi^2 + \left(\partial_{\tau} \phi \right)^2 + c^2 \left( \vec \nabla \phi \right)^2    + \frac{ \lambda_{\phi}}{4 !} \phi^4  \nonumber \\
S_{\psi,\phi} &=& \int \frac{d^{d+1} k d^{d+1} q}{\left(2 \pi \right)^{2(d+1)}}g( k, q) \bar \psi( k) \psi( k+q) \phi(q),
\end{eqnarray}
where repeated spin indices are summed.   The first term, $\mathcal L_{\psi}$,  represents a Landau Fermi liquid, with weak residual self-interactions incorporated in  forward and BCS scattering amplitudes.  The second term represents an interacting scalar boson field with speed $c$ and mass $m_{\phi}$ (which corresponds to the inverse correlation length that vanishes as the system is tuned to the quantum critical point).  The third term is the Yukawa coupling between the fermion and boson fields and is more naturally described in momentum space.  The quantity $g(k,q)$ is a generic {\it coupling function} that depends both on the  fermion momentum $\bm k$, as well as the momentum transfer $\bm q$ (we have suppressed spin indices for clarity).  For a spherically symmetric Fermi system, the angular dependence of $g(k,q)$ for $\vert \bm k \vert = k_F$ can be decomposed into distinct angular momentum channels, each of which marks a different broken symmetry.  Familiar examples include ferromagnetism (angular momentum zero) and nematic order (angular momentum $2$).   More generally, the coupling can be labelled by the irreducible representation of the crystal point group and it respects symmetry transformations under which $\phi$ and $\bar \psi \psi$ both change sign.  
 The effective action in Eq. \ref{action} will be the point of departure of our analysis below.  

\begin{figure}
\begin{center}
\includegraphics[width=0.48\textwidth]{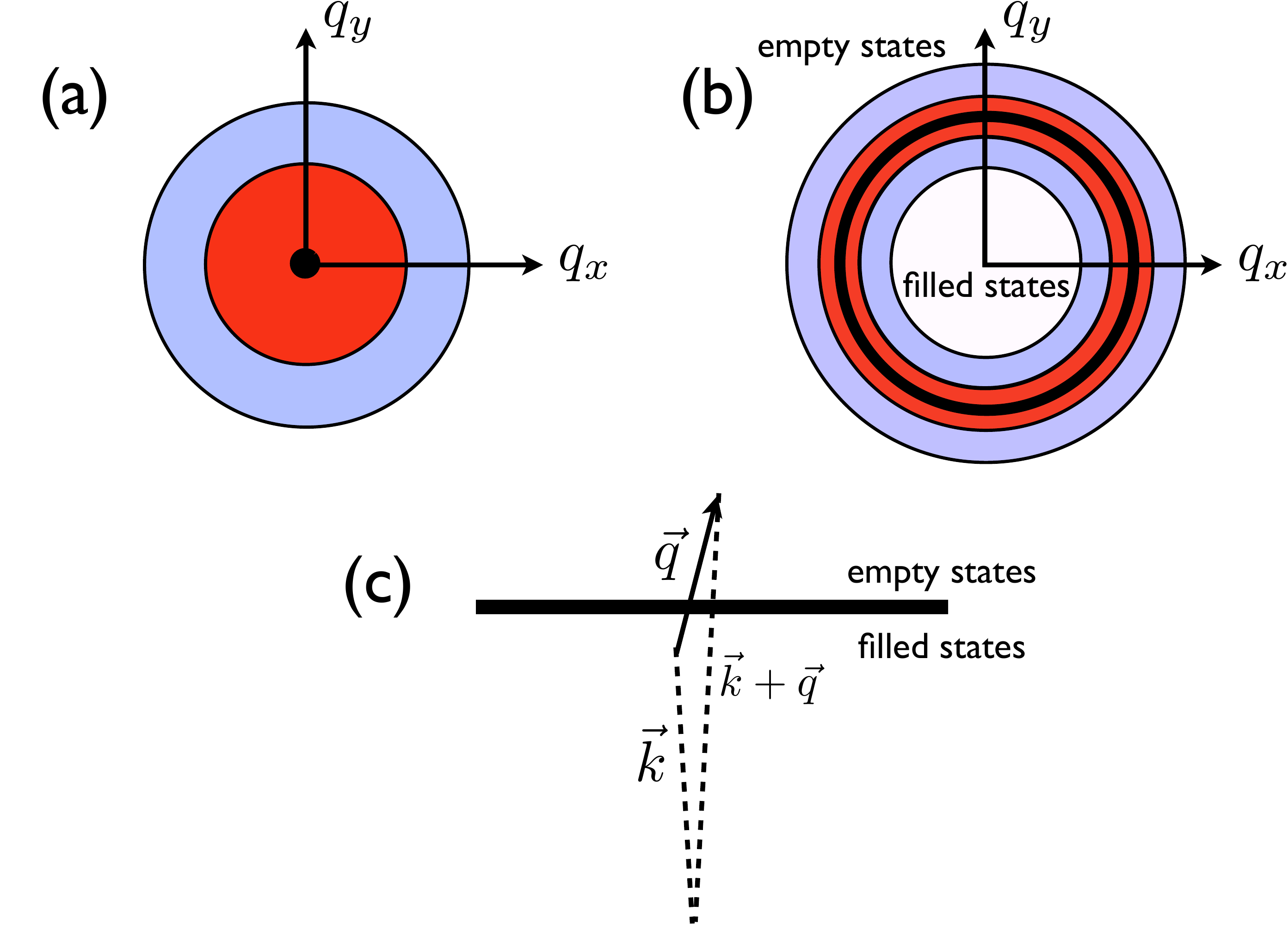}
\end{center}
\caption{Summary of tree-level scaling.  High energy modes (blue) are integrated out at tree level and remaining low energy modes (red) are rescaled so as to preserve the boson and fermion kinetic terms.  The boson modes (a) have the low energy locus at a point whereas the fermion modes (b) have their low energy locus on the Fermi surface.  The most relevant Yukawa coupling (c) connects particle-hole states  separated by small momenta near the Fermi surface; all other couplings are irrelevant under the scaling.   
}
\label{fig:scaling}
\end{figure}

We first describe a consistent scaling procedure for the action in Eq. \ref{action}.  The key challenge stems from the fact that the boson and fermion fields have vastly different kinematics. 
Our bosons have dispersion relation $k_0^2 = c^2 \bm k^2 + m_{\phi}^2$,  so that low energies correspond as usual to low momentum, and their scaling is that of a standard relativistic field theory where all components of momentum scale the same way as $k_0$. By contrast, the fermion dispersion relation is $k_0 = \epsilon(\bm k)-\mu$, so their low energy states occur close to the Fermi surface (Fig. \ref{fig:scaling}).  Moreover, the Yukawa coupling between the two sets of fields must conserve energy and momentum in a coarse-graining procedure.  These complications are easily circumvented by requiring tree-level scaling to reproduce the behavior of a Landau Fermi  liquid and a nearly-free boson decoupled from one another when $g=0$.  Furthermore, when $m_{\phi}$ is finite, we must recover Landau Fermi liquid theory: this simple notion leads to a unique scaling procedure.  As the fields are coarse grained, only the most relevant components of the Yukawa coupling function are retained.  The four fermion interaction $\lambda_\psi$ is generally also a coupling function depending on the relative orientation of the fermion momenta, with different scalings for different configurations\cite{Polchinski1992, Shankar1994}.

To be more explicit, we consider a rotationally invariant Fermi surface, and following Polchinski\cite{Polchinski1992}, we define a fermion momentum $\bm k = \bm k_F + \bm \ell$, 
where $\bm k_F$ is a point on the Fermi surface that is closest to $\bm k$; thus, $\bm \ell$ is a perpendicular displacement from the Fermi surface to $\bm k$.  As the cutoff is lowered, energies and momenta must be rescaled, and in the Fermi liquid theory, only $\bm \ell$ are rescaled while $\bm k_F$ remain unaffected.  For the boson fields, by contrast, all momenta components and energy must be rescaled as the cutoff is lowered.  We integrate out modes at tree-level with energy $\Lambda e^{-t} < E < \Lambda$, and rescale frequencies (denoted $k_0$) and momenta so that the dispersion relations remain invariant.  To simplify the discussion of scaling, we will focus on a spherically symmetric Fermi surface $\epsilon(\bm k) = \frac{1}{2 m} k^2$, so that our decomposition of the fermion momentum is equivalent to parameterizing momenta by a direction $\hat{\Omega}$ and a perpendicular magnitude $\ell$:
\be
\bm k = \bm \hat{\Omega}(k_F + \ell).
\label{eq:fermionmomentumparm}
\ee
  The dispersion relation for  
 $\ell \ll k_F$ is then simply $k_0 \approx v_F \ell$, $v_F= k_F /m$.  The natural fermion scaling is therefore to scale $\ell$ the same as $k_0$, but not to scale any other components of momentum.  In this parameterization (\ref{eq:fermionmomentumparm}), the components of momentum parallel to the Fermi surface are more properly thought of as angles rather than momenta.  
 We therefore find it natural to think of  the Fermi surface as a continuous collection of effectively $(1+1)$-dimensional fermions coupled by forward scattering and BCS interactions, as is true in an ordinary Landau Fermi liquid.

We therefore obtain the following scalings
\begin{equation} \label{eq:fermionMomentumScaling}
k_0' = e^{t} k_0,  \   \bm k_F' = \bm k_F, \ \ell' = e^{t} \ell 
\end{equation}
for the fermion states, whereas 
\begin{equation} \label{eq:BosonMomentumScaling}
k_0' = e^{t} k_0, \ \bm k' = e^{t} \bm k
\end{equation}
is the scaling that we adopt for the boson fields.  This particular scaling reflects the fact that our boson has dynamical critical exponent $z = 1$ at tree-level since we have not integrated out gapless fermions to generate a Landau-damped boson.  The fields are rescaled so that the boson and fermion kinetic energies remain invariant, which leads to the following scaling relations:
\begin{equation}  \label{eq:FieldScaling}
\psi' = e^{-3t/2} \psi , \ 
\phi' = e^{-\frac{(d+3)}{2}t} \phi
\end{equation}

From this it follows that a generic fermion interaction is irrelevant, whereas forward scattering and BCS interactions always remain marginal at tree-level: $ \lambda_{\psi}' = \lambda_{\psi}$ for all $d > 0$.  It also follows from these considerations that the boson interactions must be rescaled as 
\begin{equation}
\lambda_{\phi}'  = e^{(3-d)t} \lambda_{\phi}
\end{equation}
which sets 
$d=3$ as the upper-critical dimension for the boson fields: thus, when $g=0$ the quantum critical point has the properties of a classical critical point in one higher dimension, as is required when $z=1$.    

At first sight, scaling the momenta of the fermions differently from those of the bosons may alarm the reader.  It implies, among other things, that scale transformations in position space are non-local.  However, this feature is present even in ordinary Landau Fermi liquid theory:  the scaling procedure 
couples fermions at different points in space.  To see this explicitly, one can simply Fourier transform the momentum space scaling
  \be
  \psi(\hat{\Omega}, \ell) \rightarrow \psi'(\hat{\Omega}, \ell) = e^{-3 t/2} \psi(\hat{\Omega}, e^{-t } \ell),
  \ee
back to position space.  At linear order in $t$, one finds that $\psi'(x)$ depends on an integral over all $\psi(x)$. 
  We therefore are led to study the scaling of all couplings in momentum space.   Finally, we note that away from criticality, when $m_{\phi} \ne 0$, the boson can formally be integrated out and we must recover Landau Fermi liquid theory: simplicity demands that the field scaling  should not depend on $m_{\phi}$, which further compels us to adopt this scaling procedure.

Next, we consider the fate of a non-zero Yukawa coupling under this scaling procedure.  In the low energy limit, the Yukawa coupling function $g(k,q)$ has a Taylor expansion of the form
\begin{equation}
g(k,q) = g(\bm k_F,0) +a_1 \ell  + a_2 q + \cdots 
\end{equation}
and it follows that under the scaling procedure above, only $g(\bm k_F,0)$ is marginal whereas all other terms are irrelevant.   
Thus, only small momentum transfers imparted by the boson remain marginal in $d=3$ as we scale $k$ towards the Fermi surface (see Fig. \ref{fig:scaling}):
\begin{equation}
\lim_{  \bm k \rightarrow \bm k_F} \lim_{q \rightarrow 0} g'(\bm k', \bm q') = e^{\frac{3-d}{2} t} g(\bm k, \bm q) 
\end{equation}
This simple relation is derived explicitly in the Appendix.  
We shall refer to the coupling $g(\bm k_F, 0)$ simply as $g$ for the remainder of the paper.  We see that $d=3$ is the upper-critical dimension for the Euclidean action; below it, both $\lambda_{\phi}$ and $g$ are relevant.  Thus, we can naturally expect to find new fixed points in an $\epsilon-$expansion, which we show in the next section.

  Some readers may be familiar with other scaling schemes, such as the `patch' picture, where all components of the fermion momenta are scaled towards a single point on the Fermi surface.  In this scheme, the fermion dispersion relation takes the form $k_0 =  v_F k_\perp + \bm k_\parallel^2/2m$, and so the fermions scale with both $k_\perp$ and $\bm k_\parallel$.  However, this scheme cannot be applied to the entire smooth Fermi surface without breaking it up into patches in an arbitrary way, with an increasing number of patches needed as we evolve to lower energies.  
 Most importantly, the ``patch" scaling approach has the unappealing feature that forward scattering and BCS interactions are irrelevant (see appendix).  This leads to the apparent contradiction that when the system is tuned away from the critical point, Fermi liquid behavior is not recovered.  This can only be fixed by resorting to a more complicated procedure\cite{Lee2008}.

In Hertz's approach, the boson kinetic term contains a non-analytic  self-energy correction that one obtains upon integrating out gapless fermions on the Fermi surface:
\begin{equation}
\label{landaudamping}
S_{\phi}^{\rm Hertz} = S_{\phi} +g^2 m_{\psi}^2 \int  \frac{d^{d+1}q}{\left( 2 \pi \right)^d} \frac{\vert q_0 \vert}{\vert q \vert } \theta(\vert q \vert  - \vert q_0 \vert)  \phi_{\bm q} \phi_{-\bm q}  
\end{equation}
The inclusion of this term in the bare action reduces the upper-critical dimension of the boson fields  by 3: thus for $d> 1$, the bosons are imagined to be described by their gaussian fixed point.  In this case, the scaling of time and space is different for bosons and fermions\cite{Sachdev}.  
By contrast, in our theory, the self-energy correction in Eq. \ref{landaudamping}  {\it does not } occur in our starting action, since we have integrated out only the high energy modes.  

It is important to stress that although the physics of Landau damping is not incorporated directly into our bare action, it is always present in the theory: it is clear from Eq. \ref{action} that we would reproduce Landau damping when we integrate out fermions.  In other words, the Landau damping effect is a property of the low-energy correlators of our theory at weak coupling.  However, it does not alter the scaling of the theory at energies that are large compared to $g m_\psi$, which will be a parametrically small scale throughout our analysis.   By choosing to keep the low energy fermions, we show that we obtain an entirely new description of a non-Fermi liquid metal.

\section{Fixed point structure at one-loop}

In this section, we discuss the fixed-point structure of the theory (\ref{action}), setting
the four-Fermi interaction $\lambda_{\psi}=0$ for now.  This is common also in treatments of
Fermi liquid theory, where one first finds the Fermi liquid fixed point, and then assesses its
stability to fermion self-interactions.  We discuss the non-trivial effects of four-Fermi interactions
in \S4 and \S5.

\begin{figure}[h!]
\begin{center}
\includegraphics[width=0.48\textwidth]{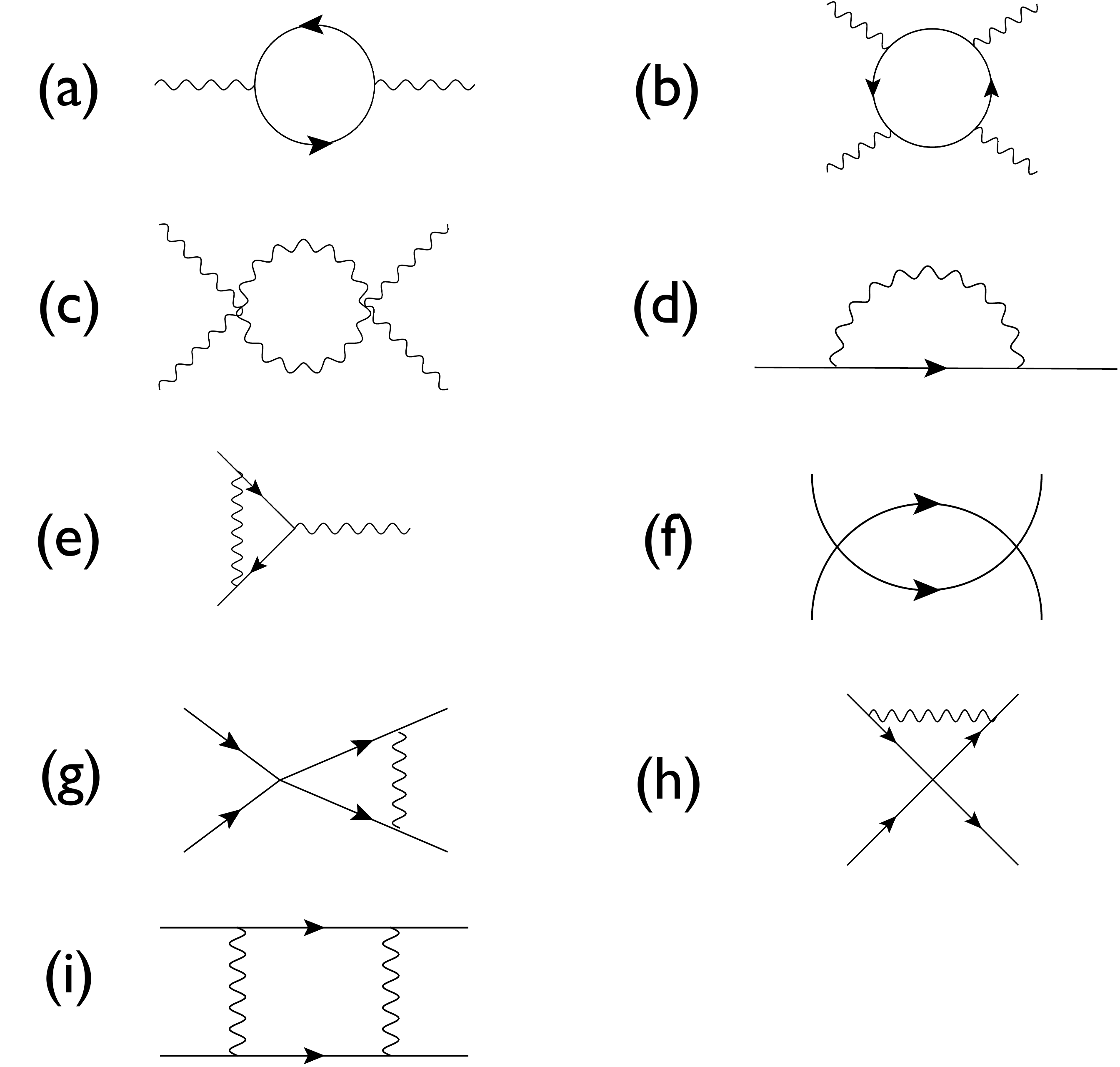}
\end{center}
\caption{One-loop diagrams.  The boson self-energy (a), boson self-interactions (b,c),  fermion self-energy (d), vertex correction (e) and particle-hole scattering (f). Diagrams (a) and (b) do not contribute to the renormalization group flow while (c) produces the ordinary Wilson-Fisher fixed point for bosons.  Diagram (d) gives rise to fermion wave-function renormalization and (e) yields logarithmic Yukawa coupling constant renormalization.  The usual marginal BCS interaction Fermi liquid theory (f) is altered  by fermion wave-function renormalization as well as by diagram (g), both of which make the BCS interaction irrelevant.    
 }
\label{fig:alldiagrams}
\end{figure}

Since both $g$ and $\lambda_{\phi}$ are relevant below $d=3$, non-trivial fixed points can be obtained in a systematic expansion in $\epsilon = 3-d$.   In this section we describe the  renormalization group flow to one-loop order that is obtained when the diagrams in Fig.  \ref{fig:alldiagrams} are taken into account.  In all of these diagrams, internal propagators have energies in an infinitessimal  shell  $ \Lambda e^{-t} < E < \Lambda$, whereas external legs have energy $E < \Lambda e^{-t}$.  The leading contribution from these diagrams is obtained by performing the loop integrals in $d=3$.  After  mode elimination, we rescale energy, momenta and fields in order to preserve the boson and fermion kinetic terms.    We first summarize the effect of each of the one-loop diagrams in Fig. \ref{fig:alldiagrams};  explicit derivations can be found in the appendix.  

The first two diagrams (Fig. \ref{fig:alldiagrams}(a) and (b)), represent the fermion contribution to the boson self-energy and self-interactions respectively.   Curiously, neither of them contribute to the RG flow.  The boson self-energy obtained from eliminating fermion modes in the shell is proportional to 
\begin{equation}
\Pi_{d \Lambda}(p) \propto \int_{d \Lambda} d q_0 \left[ {\rm sgn}(q_0) - {\rm sgn}(q_0 + p_0) \right] = 0,
\end{equation}
 since the external frequencies are by definition smaller than those in the shell being eliminated.  A similar result is obtained for the diagram in Fig. \ref{fig:alldiagrams}(b).  Therefore, the fermions do not affect the running of  $\lambda_{\phi}$ to one-loop order, and there is no boson wave-function renormalization to one-loop order.  
 
 The diagram in Fig.\ref{fig:alldiagrams}(c) yields the standard $\mathcal O(\lambda_{\phi}^2)$ contribution:
 \begin{equation}
 \label{wilson_fisher_flow}
 \frac{d \lambda_{\phi}}{d t} = \epsilon \lambda_{\phi}  - a_{\lambda_{\phi}} \lambda_{\phi}^2.
 \end{equation}
 Here, $a_{\lambda_{\phi}} $ is a positive constant given in the appendix and  $t=-\log \left[ \Lambda/ \Lambda_0 \right]$ is now the RG flow parameter.

Next, consider the fermion self-energy in Fig \ref{fig:alldiagrams}(d).  This diagram determines the fermion wavefunction renormalization, and therefore affects the running of all fermion couplings.  It produces a contribution of the form
\begin{equation}
\Sigma_{d \Lambda}(k) =  -  \left( i k_0 g^2 a_g \right)  d \log{\Lambda} , 
\end{equation}
where $a_g$ is a constant and is derived in the appendix.

The first contribution to the flow of the Yukawa coupling comes from fermion wave-function renormalization.  
The second contribution to the flow of $g$ comes from the vertex correction diagram in Fig.  \ref{fig:alldiagrams}(e) with zero external boson momentum.  This quantity is related to the fermion self-energy via the simple relation
\begin{equation}
\frac{\delta g_{d \Lambda}(k, 0)}{g} =   C_3 \frac{ \partial \Sigma_{d \Lambda}(k) }{\partial \left( i k_0 \right) }  = -g^2 C_3 a_g d \log{\Lambda}
\end{equation}  
where $C_3$ is a constant which is equal to 1 in the theory with a single scalar field $\phi$, and which takes more general values in large N theories we discuss later in the paper.

After the mode elimination and field rescaling, it can immediately be seen that the Yukawa coupling becomes 
\begin{equation}
g'(t) = \left( g - C_3 a_g g^3 t \right) e^{-g^2 a_g t}e^{\epsilon t/2},
\end{equation} 
where the first factor incorporates the vertex correction, the second takes into account wave-function renormalization, and the exponential factor is obtained from scaling at tree level.  The resulting flow equation for the Yukawa coupling is readily obtained:
\begin{equation}
\label{yukawa_flow}
\frac{ d g }{d t} = \frac{\epsilon}{2} g - (1+C_3) g^3 a_g + \mathcal O(g^3 \epsilon)
\end{equation}
Note the existence of a fixed point at $g^2 = \frac{\epsilon}{2(1+C_3) a_g}$.

Equations \ref{wilson_fisher_flow} and \ref{yukawa_flow}  are the key results of this section.  To one-loop order in the $\epsilon$-expansion, there is a non-trivial fixed point which has 
2 main features.  Firstly, the boson flows to the usual Wilson-Fisher fixed point with $\lambda_{\phi}^* = \mathcal O(\epsilon)$, and is surprisingly unaffected by the Fermi surface to this order.  Secondly, there is a non-trivial fixed point at finite $g^* = \mathcal O(\sqrt{ \epsilon})$ which corresponds to a non-Fermi liquid in which the fermion propagator has an anomalous dimension.  

When $\epsilon \ll 1$, the fixed point values  $\lambda^*_{\phi}$ and $g^*$ are small.  Therefore, the properties of the system can be computed via perturbation theory.  Correlation functions of the boson conform with the predictions of the Wilson-Fisher fixed point to one-loop order as do the critical exponents, whereas the fermion propagator develop branch cuts signifying the loss of a well-defined quasiparticle.  To be more explicit, we compute the anomalous dimension $\gamma_{\psi}$ of the fermion propagator directly from the expression for the $\beta$ function and the RG equation: 
\begin{equation}
\beta_{g^*} = \left[ \frac{3-d}{2} - 4 \gamma_{\psi} \right] g^*  = 0
\end{equation}
which includes the contribution $C_3 = 1$ from the vertex correction, from which we see that $\gamma_{\psi}  = \epsilon/8$.  The fermion propagator at the fixed point has the scale-invariant form\cite{Phillips2013}
\begin{equation}
G(\omega, \ell) \sim \frac{1}{(i \omega - v_F \ell)^{1-2 \gamma_{\psi}} } f(\omega/\ell),
\end{equation}
where $f(\omega/\ell)$ is a scaling function that is undetermined by the RG equation.
Note that this scale invariant form, which follows from the existence of a fixed point, is obtained despite the fact that the fermions are at finite chemical potential.  
Therefore, at the fixed point,  
the imaginary part of the  fermion self-energy varies as 
\begin{equation}
\label{imsigma}
{\rm Im} \Sigma(k) \sim (g^*)^2 \omega^{1 - \frac{\epsilon}{4}}.  
\end{equation} 
By contrast, recent predictions based on the standard approach to the problem\cite{Sachdev} suggest that
\begin{equation}
{\rm Im} \Sigma(k) \sim \omega^{d/3} =  \omega^{1 - \frac{\epsilon}{3}}.  
\end{equation} 
where the second equality makes the comparison directly with the expression in Eq. \ref{imsigma} obtained at the fixed point.  We note that even to leading order in $\epsilon$, there are discrepancies between the two sets of theories.

\section{fermion Self-Couplings}
\label{sec:fermionselfcoupling}

\begin{figure}
\begin{center}
\includegraphics[width=0.48\textwidth]{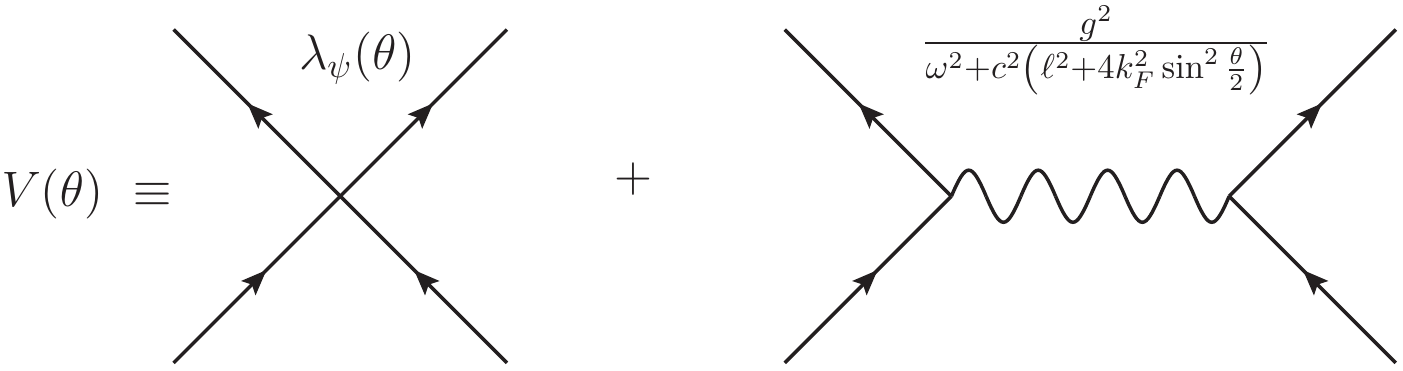}
\end{center}
\caption{ This figure depicts the sum of two diagrams that define the tree-level fermion scattering function $V(\theta)$.  This function then appears directly in the generalized BCS loop diagrams depicted in Fig.~\ref{fig:alldiagrams}(f, g, i).
 }
\label{fig:VDiagram}
\end{figure}

Next, we describe the fate of four fermion interactions at the non-trivial fixed point.  In addition to the Fermi liquid contribution in Fig. \ref{fig:alldiagrams}(f), there are several diagrams which affect the  flow of $\lambda_{\psi}$, and some of their contributions are rather subtle.  

Aside from the fermion wave-function renormalization of diagram Fig. \ref{fig:alldiagrams}(d), there are four diagrams that produce vertex corrections. 
The simplest vertex correction to consider is depicted in Fig. \ref{fig:alldiagrams}(h).
As explained in the appendix, due to the kinematic constraints of the Fermi surface, it has no divergent piece.

The other vertex corrections shown in Fig. \ref{fig:alldiagrams}(g) and Fig. \ref{fig:alldiagrams}(i) 
are a bit more subtle.  For example, if one directly evaluates any of these diagrams individually one seems to find a dependence of the form $\log^2( \Lambda )$, which would naively lead to explicit $\Lambda$ dependence in the RG flow equations.  

One may obtain more insight into the origin of these puzzling divergences by slightly re-organizing the calculation.  For this purpose, let us define the general vertex function 
\be
V(\theta, \omega) \equiv \lambda_\psi(\theta) + \frac{g^2}{\omega^2 + c^2 ( \ell^2 + 4k_F^2 \sin^2 \frac{\theta}{2})}
\ee
where $\theta$ is the angle between the initial and final antipodal fermions, and we have suppressed the explicit $\ell$ dependence of $V$ as well as the scale dependence of the couplings.  
Inserting this vertex function into one-loop Feynman diagrams, one can see that in the integral over momenta, soft scattering at small $\theta$ is responsible for an enhanced log divergence on top of the one which follows from naive scaling.

 One simple way of seeing the enhanced log divergences is as follows.  We can write all of the diagrams Fig. \ref{fig:alldiagrams}(f, g, i) in the compact form
\be \label{eq:GeneralizedBCSAngular}
I(\Omega) =  \int \frac{\delta \Lambda d \ell \, d^{d-2} \Omega' \, k_F^{d-2}}{(2 \pi)^d (\Lambda^2 + v_F^2 \ell^2)}  V(\Omega', \Lambda) V(\Omega - \Omega', \Lambda) \ \ \
\ee
Now we see that the integral over $\Omega'$ is merely a convolution, so we can diagonalize it by writing the $V(\theta, \omega)$ in terms of spherical harmonics on the Fermi surface, giving
\be \label{eq:GeneralizedBCSLoop}
I(L) = a_{V}  V(L, \Lambda)^2 d \log (\Lambda)
\ee
for a constant $a_V$.  The $\log^2(\Lambda)$ effect alluded to above is now contained in the behavior of $V_L$ as a function of Fermi surface angular momentum $L$.

Similarly log$^2$ divergences have been encountered in the earlier literature
\cite{Nayak1994a,Son1999,Schafer1999,Schafer2001,Wang2013}.  An interesting suggestion for an improved RG to surmount this issue appears in the work of Son\cite{Son1999}.  We will see here that in controlled large $N$ limits, the log$^2$ diagrams do not contribute, and the RG analysis is conventional.  We intend to pursue a systematic investigation of the proper treatment of such divergences at small $N$ in further work.

\section{One-Loop Structure at Large N and fixed points including $\lambda_{\psi}$}

In \S3, we discovered new one-loop fixed points where a Wilson-Fisher boson dressed a Fermi liquid into a non-Fermi liquid.  However, incorporation of four-Fermi interactions in \S4\ leads to new issues in the renormalization group, whose systematic investigation we leave for the future.
For now, we can gain significant theoretical control over the model including four-Fermi interactions by introducing a large $N$ version of it.  

To do this, the fermions $\psi$ are promoted to $N$-vectors $\psi_i$ while the scalar is promoted to an $N\times N$ complex matrix $\phi^i_j$:
\begin{eqnarray}
\mathcal L_{\psi} &=&   \bar \psi^i \left[ \partial_{\tau}  + \mu -\epsilon(i  \nabla)  \right] \psi_{i} 
+ \frac{\lambda_{\psi} }{N}\bar \psi^i \psi_{i} \bar \psi^j \psi_{j} 
\nonumber  \\
\mathcal L_{\phi} &=& {\rm tr}\left( m_{\phi}^2   \phi^2 + \left(\partial_{\tau} \phi \right)^2 + c^2 \left( \vec \nabla \phi \right)^2\right)   \nonumber \\
&&+  \frac{\lambda_\phi^{(1)}}{8 N} \tr (\phi^4) + \frac{\lambda_\phi^{(2)}}{8 N^2} (\tr (\phi^2))^2  \nonumber \\
\mathcal L_{\psi,\phi}&=& \frac{g}{\sqrt{N}}   \bar \psi^i \psi_{j} \phi^j_i
\label{largeNaction}
\end{eqnarray}
where we consider spin-less fermions for simplicity. In order to avoid having more than one independent scalar mass that must be tuned to zero at the fixed point, we take $\phi^i_j$ to be in an irreducible representation.  For concreteness, we take this to be the adjoint of $SU(N)$, and the fermions to transform in the fundamental representation, though the leading large $N$ effects will not be especially dependent on this choice. 

Although a priori there are two distinct quartic boson couplings with different trace structures, one can in fact set
$\lambda_{\phi}^{(1)} = 0$ in a natural way.  The model enjoys an enhanced $SO(N^2)$ symmetry in that
limit (softly broken by the relevant parameter $g$), and so it is radiatively stable to do so.  We proceed with $\lambda_{\phi}^{(1)} = 0$.

In appendix \ref{app:matrix}, we evaluate the generalization of the renormalization coefficients $a_g, a_{\lambda_\psi}, a_V$, and $C_3$ to large $N$, but here we will just give a qualitative overview.  First of all, the diagram  
(Fig. \ref{fig:alldiagrams}(a)) vanishes in this limit,  including the finite piece.  This implies that Landau damping is completely absent at $N \rightarrow \infty$. 
This large $N$ limit in $d=3$ was exploited previously in \onlinecite{Mahajan2013}.
This is interesting, but it does mean that the large N limit and the $\omega \to 0$ limit (where Landau
damping is dominant in other treatments of this problem) exhibit subtle interplay.

In fact, this is part of a more general point: at infinite $N$, the fermions do not renormalize the scalar at all.  The fixed point properties of the scalars can therefore be studied independently of the fermions.  
On the other hand, the scalars do have an effect on the renormalization of the fermions, even though the fermions do not ``back react'' on the scalars.  In particular, the wavefunction renormalization, which affects the running of all fermion interactions, survives at leading order in large $N$:
\be
a_g = \CO(N^0).
\ee
In contrast, the direct cubic vertex renormalization diagram Fig. \ref{fig:alldiagrams}(e)  vanishes at large $N$:
\be
C_3 &=& \CO(N^{-2}).
\ee
The factor of $1/N$ in the four-fermion interaction $\lambda_\psi$ has been chosen so that the resulting one-loop renormalization of the fermion propagator is finite at $N\rightarrow \infty$.  The interaction itself naively has two different possible structures, $\bar{\psi}^i(p)\psi_i(p')\bar{\psi}^j(-p)\psi_j(p')$ and $\bar{\psi}^i(p)\psi_j(p')\bar{\psi}^i(-p)\psi_j(p')$ in the action above. However, by anti-commuting the fermions and relabeling the momenta, these can be seen to be just a single interaction (with a coefficient $\lambda_\psi$ that can be a function of the angle $\cos \theta = \hat{p} \cdot \hat{p}'$). 
The diagrams in Fig. \ref{fig:alldiagrams}(g), (h), and (i) do not contribute to $\lambda_\psi$ at leading order in large $N$.
Consequently, the double-logs mentioned in section \ref{sec:fermionselfcoupling} are also absent at infinite $N$. \footnote{A different class of diagrams contributing to the overlap region between forward and antipodal scattering is not suppressed at large $N$, and was considered in [\onlinecite{SonShuster}].  The interesting effects of these diagrams set in at a scale that is exponentially small at small $\epsilon$, and we will not discuss them further.}    

The structure of the resulting RG fixed points is as follows.  The boson still behaves as if it is at a Wilson-Fisher fixed point at leading order in large N.  The fermion is dressed into a non-Fermi liquid, and we can now assess the stability of this non-Fermi liquid fixed point to superconducting instabilities.  The leading RG equation for $\lambda_{\psi}$ is therefore of the form
\begin{equation}
\label{nosuper}
{d\over dt} \lambda_{\psi} = -2a_g g^2 \lambda_{\psi}.
\end{equation}  An additional term on the right-hand side of the form $\beta \lambda_\psi^2$  (i.e. the conventional BCS result, 
that would drive attractive interactions to grow at low-energy) is absent at infinite $N$; more precisely, $\beta = \CO(N^{-1})$.  We therefore see that 
there is a ${\it stable}$ fixed point at $\lambda_{\psi}=0$.  The large $N$ fixed point is stable against superconductivity.   Similar conclusions were obtained in a recent study\cite{Chung2013}  of superconductivity of fermions at finite density coupled to U$(1)$ gauge fields in $d=3+1$.  

\section{Discussion}

The key result of this paper is the existence of a non-Fermi liquid fixed point for a metal near a quantum critical point below $d=3$.  We have obtained the non-Fermi liquid fixed point in a theory which formally integrates out only high energy modes, never incorporates the effect of Landau damping in the tree-level action, and treats the low energy boson and fermion modes on an equal footing.  Furthermore, in the large-N limit of \S5,  the non-Fermi liquid is always stable against superconductivity, since BCS interactions become irrelevant at the fixed point due to the first term of Eq. (\ref{nosuper}), which is $\mathcal O(\epsilon)$.  

However, there is a peculiar aspect of the fixed points obtained in the present analysis.  We have emphasized that while the physics of Landau damping can always be recovered in our theory at any stage by integrating out fermions, there is no sense in which Landau damping smoothly grows under the RG.  The reason for this effect is that the boson self-energy to one-loop order is not logarithmically divergent.  Stated differently, the damping only arises when fermion modes below the boson energy are integrated out; in our Wilsonian RG treatment, we never integrate out these modes.  It is a curiosity that in our approach, the boson dynamical critical exponent remains at unity at all finite steps of the RG.  It seems very likely that as one goes to higher orders in $\epsilon$, more complicated diagrams generate
a non-trivial dynamical critical exponent for the scalar, and its behavior no longer coincides with that
of a scalar at the Wilson-Fisher fixed point.  At $\epsilon=1$, such corrections are likely to be
quite important.

This raises the natural question of whether the fixed points we found here can be continuously interpolated to $\epsilon=1$, and govern the IR behavior of a quantum critical metal in $d=2+1$. While there is good reason to believe that the fixed point structure survives to $\epsilon = 1$, there are several issues which present a challenge in constructing the theory.  Firstly, at energy scales below $\omega_{LD} \sim g m_{\psi}^*$, the boson becomes substantially Landau damped due to the presence of the dissipative fermion bath\cite{Mahajan2013}.    Secondly, the effect of $\lambda_{\phi}$, which is $\mathcal O(1)$ relevant in $d=2+1$, should be taken into account beyond one-loop order.  A promising route to describing the non-Fermi liquid in $d=2+1$ involves looking for self-consistent solutions of both the boson and fermion self-energies by computing a Dyson expansion for both quantities.  It is well-known that starting with a Fermi liquid propagator, self-consistency cannot be achieved, since $g$ is $\mathcal O(1)$ relevant.  Stated differently, the boson Landau damping, which assumes an underlying Fermi liquid, leads to a non-Fermi liquid description in $d=2+1$.  However, it is conceivable that by starting with a non-Fermi liquid ansatz, a self-consistent solution to the problem at hand may be achieved.  In suitable large $N$ limits, for instance, one can find closed-form integral equations for the boson and fermion self-energies, and search for self-consistent solutions.  Results in this direction, with a comparison to the present work, will be described in a future publication.

\acknowledgements{We acknowledge important conversations with S. Chakravarty, A. Chubukov, E. Fradkin, S. Hartnoll, S. Kivelson, S.-S. Lee, R. Mahajan, M. Metlitski, M. Mulligan, and S. Sachdev. This work was supported in part by the National Science Foundation grant PHY-0756174 (SK), DOE Office of Basic Energy Sciences, contract DE-AC02-76SF00515 (SK and SR), the John Templeton Foundation (SK and SR), and the Alfred P. Sloan Foundation (SR).  This material is based upon work supported in part by the National Science Foundation Grant No. 1066293. ALF and JK were partially supported by ERC grant BSMOXFORD no. 228169. JK acknowledges support from the US DOE under contract no. DE-AC02-76SF00515.
}

\newpage

\appendix
\begin{widetext}

\section{Calculational Details}

\subsection{Scaling dimension of the Yukawa coupling function}

 We would like to study the scaling of the  Yukawa coupling function, which can be written with full detail as
\begin{equation}
S_{\psi,\phi} = \int \frac {d^{d+1}k d^{d+1} k' d^{d+1}q}{\left( 2 \pi \right)^{3(d+1)}} g(k,q) \bar \psi(k') \psi(k) \phi(q) \delta^{(d)}( \bm k' - \bm k - \bm q) \delta(k'_{0} - k_{0} - q_{0})
\end{equation}
where $k_1,k_2$ label the fermions, and $q$ labels the bosons.  We have written, {\it e.g.}, $k = (k_0, \hat k_F + \vec \ell \, )$ for the fermions with  $\vec \ell$ perpendicular to the Fermi surface. Note that we have included explicitly the energy and momentum conserving $\delta$-function, which plays an important role in the scaling analysis.  The coupling function has an expansion of the form
\begin{equation}
g(k,q) = g(\bm k_F,0) + a_1 \ell + a_2 q + \cdots
\end{equation}
We next perform the tree-level scaling analysis described in Section II, paying careful attention to the behavior of the momentum conserving $\delta$-function, which is written more explicitly as
\begin{equation}
\delta^{(d)} \left( \bm k' - \bm k - \bm q \right) = \delta^{(d)}(\bm k_F'  - \bm k_F   + \bm \ell' - \bm \ell - \bm q) 
\end{equation}
Following the analysis of Polchinski\cite{Polchinski1992}, we consider two cases: 
\begin{equation}
{\bf Case \ I}:  \ell, \ell' \ll \vert \bm k_F' - \bm k_F \vert 
\end{equation}
and
\begin{equation}
{\bf Case \ II}:  \bm k_F \approx \bm k_F'
\end{equation}
The momentum $\delta$-function scales only in the second case and has scaling dimension $-1$, {\it not} $-d$.  The energy conserving $\delta$-function always scales.  Under the scaling transformation, the various components of the Yukawa coupling function scale as
\begin{eqnarray}
g(\bm k_F,0) &\rightarrow& g(\bm k_F,0) \nonumber \\
a_1 \ell  &\rightarrow & a_1 \ell e^{-t} \nonumber \\
a_2 q &\rightarrow & a_2 q e^{-t} 
\end{eqnarray}
whereas the combination of scaling the measure and fields produces a factor $e^{\frac{1-d}{2}t}$.  We conclude, therefore, that $a_1, a_2$ are irrelevant.  Retaining only $g(\bm k_F,0)$, the Yukawa coupling scales as
\begin{eqnarray}
{\bf Case \ I}: \ \ \ \ \   S_{\psi, \phi} \rightarrow e^{\frac{1-d}{2}t} S_{\psi, \phi}   \nonumber \\
{\bf Case \ II}: \ \ \ \ \   S_{\psi, \phi} \rightarrow e^{\frac{3-d}{2}t} S_{\psi, \phi} 
\end{eqnarray}
The difference between the two cases arises because of the scaling of the $\delta$-function in Case II.  Thus we conclude that 
\begin{equation}
\lim_{  \bm k \rightarrow \bm k_F} \lim_{q \rightarrow 0} g'(\bm k', \bm q') = e^{\frac{3-d}{2} t} g(\bm k, \bm q) 
\end{equation}
so that for small boson momentum, and for fermion states close to the Fermi surface, the Yukawa coupling is relevant below $d=3$.  Note that the Yukawa coupling $g(\bm k_F,0)$ can exhibit non-trivial angular dependence on the Fermi surface and still remain marginal in $d=3$.

\subsection{Shortcomings of the ``patch" scaling procedure}
For completeness, we show here that in the ``patch" scaling scheme, the BCS four fermion couplings are irrelevant in all $d>1$
  To see this, start with the quadratic action written as
  \be
  \int d k_0 d \ell d^{d-1} k_\parallel \psi (\partial_\tau - v_F \ell - k_\parallel^2/2m) \psi.
  \ee
  Scaling $[k_0]=[\ell]=2[k_\parallel]$, in order to keep the action invariant one must assign the fermions a scaling of $[\psi] = - \frac{d+5}{4}$.  Passing to the four-fermion interaction, we have
  \begin{equation}
  \int \prod_{i=1}^4 \left( d k_0^{(i)} d\ell^{(i)} d^{d-1} k_\parallel^{(i)} \right) \psi_1 \psi_2 \psi_3 \psi_4 \delta^{(d+1)}\left(\sum_i k^{(i)}\right).
  \end{equation}
  When the fermions are arranged antipodally on the Fermi surface so that the large components of their momenta cancel inside the $\delta$ function, the total scaling of this interaction is $\frac{3}{2} (d-1) [k_0]$, and so is not marginal above one spatial dimension.  This is in contrast with the scaling that we adopt, for which the four-fermion interaction on antipodal fermions is classically marginal in any dimension.\cite{Polchinski1992}

\subsection{Determination of RG Flow from Divergences}

The standard Wilsonian renormalization group procedure is complicated by the kinematics of a Fermi surface.  The main issue is that very large momentum transfers can contribute to very low-energy processes.  For example, low energy fermions at generic points on the Fermi surface scatter via the exchange of bosons with Fermi-scale momentum.

This means that we must be careful to precisely specify our RG scheme.  In what follows, we will integrate out modes in frequency shells with unconstrained momenta, so that
\be
e^{-t} \Lambda < \omega < \Lambda \ \ \ \mathrm{and}  \ \ \ 0 \leq | \vec k | < \infty
\ee
Other schemes where we also decrease the cutoff on all $| \vec k|$ can be considered, but we will adopt this scheme since it is the one that we find most convenient.

 In the sections that follow we will compute the UV divergent part of various Feynman diagrams that renormalize the kinetic terms and couplings.  The $\beta$ functions for the couplings are related to these UV divergences in an elementary way. All of the relevant diagrams are shown in Fig. \ref{fig:alldiagrams}.  We treat each of these in turn.
 
\subsection*{\begin{flushleft} (a) \end{flushleft}}

This will be given by the integral over a fermion loop
\be
\Pi(p) = g^2 \int_{\Lambda(1 - \delta)}^\Lambda d \omega \int d^{d-1} \hat \Omega k_F^{d-1} d \ell \frac{1}{i \omega - v_F \ell} \frac{1}{i (\omega - i p_0) - v_F \ell -  v_F \hat \Omega \cdot  \vec p}
\ee
An equivalent contribution arises from the small window of negative frequencies $-\Lambda < \omega < -\Lambda (1-\delta)$.  We are interested in the regime where the boson energy and momentum $(p_0, \vec p)$ are much smaller than $k_F$.  This means that the boson must split into roughly antipodal fermion pairs.  The first fermion has energy $\omega$, lives at $\hat v_F$ on the fermi surface, and deviates from the fermi surface by $\ell$.  The second fermion is nearly antipodal, but has its momentum and energy fixed by momentum conservation.  We can again immediately evaluate the trivial frequency integral, giving
\be
\Pi(p) = g^2 \delta \Lambda  k_F^{d-1}  \int d^{d-1} \hat \Omega d \ell \frac{1}{i \Lambda - v_F \ell} \frac{1}{i (\Lambda - i p_0) - v_F \ell -  v_F \hat \Omega \cdot  \vec p} + (\Lambda \rightarrow - \Lambda)
\ee
The $d \ell$ integration can be performed by residues, and a non-vanishing result requires that it have poles on both sides of the real axis.  This leads us to the conclusion that $\Pi_{d \Lambda}(p)$ vanishes for $\Lambda > |p_0|$, as claimed in \S3.

\subsection*{\begin{flushleft} (b) \end{flushleft}}

The boson quartic coupling $\lambda_\phi$ is classically marginal in $d=3$ and so UV divergences are insensitive to the external momenta, which can be set to zero.  The one-loop renormalization due to a fermion loop is simply
\be
2g^4   \int \frac{ d \ell k_F^{2} d^{2}\hat{\Omega} }{(2 \pi)^{d+1}}  \frac{1}{(i \Lambda - v_F \ell)^4}.
\ee
The $ d \ell$ integral can be done by contour integration, and is then easily seen to vanish since all poles lie on one side of the real axis.  Therefore this diagram does not contribute to the running of $\lambda_\phi$.

\subsection*{\begin{flushleft} (c) \end{flushleft}}

The renormalization of $\lambda_\phi$ due to self-interactions, on the other hand, has a well-known logarithmic divergence in $d=3$.  For completeness, let us reproduce this here using our shells of restricted frequency (but unrestricted momentum):
\be
\delta \lambda_\phi &=& \left(\frac{3}{2}\right)(2) \delta \Lambda  \lambda_\phi^2 \int \frac{ d^3 p}{(2\pi)^{4}} \frac{1}{(\Lambda^2 + p^2)^2} = \frac{3 \lambda_\phi^2}{16 \pi^2} \frac{\delta \Lambda}{\Lambda}.
\ee
The pre-factor of $\frac{3}{2}$ is due to diagrammatics: there are three diagrams that contribute (s-,t-, and u-channel), each with a symmetric factor of $1/2$, and the additional pre-factor of 2 is from the two thin shells, at positive and negative frequencies.  

\subsection*{\begin{flushleft} (d) \end{flushleft}}

We start with an inflowing fermion with frequency $p_0$ and momentum $\vec{p} = \Omega(k_F + \tilde{\ell})$.  Take the internal fermion to have frequency $p_0 + \omega$ and momentum $\vec{p}' = \Omega'(k_F + \ell)$.  Then, the boson frequency is $-\omega$, and its momentum-squared is
\be
(\vec{p}- \vec{p}')^2 &=& 2 k_F^2 (1-\cos \theta) + (\ell- \tilde{\ell})^2 + (2 k_F (\ell+ \tilde{\ell}) + 2 \ell \tilde{\ell})(1-\cos \theta).
\ee
The contribution to the loop diagram from bosons with $\CO(1)$ changes in the fermion angle will be very suppressed - in fact, in order to get a non-suppressed contribution, one must have $1-\cos \theta \sim 1/k_F^2$.  This is why we separated out the last term in brackets in the expression above - its contribution is $1/k_F$-suppressed compared to the other terms in the region that contributes.  So, let us drop this term in the following.  We will then shift integration variables $\ell \rightarrow \ell+ \tilde{\ell}$, to obtain
\be
\Sigma(p_0, \tilde{\ell}) &=& g^2 \int \frac{d \omega d \ell k_F^2 d \Omega'^2}{(2\pi)^{4}} \frac{1}{\omega^2 + c^2 \ell^2 + 2 c^2 k_F^2 (1-\cos \theta) } \frac{-1}{i ( \omega + p_0) - v_F(\ell + \tilde{\ell}) }
\ee
Now, we see that we can define $m \equiv i p_0 - v_F \tilde{\ell}$, to define 
\be
\Sigma(m) &=& g^2   \int \frac{d \omega d \ell k_F^{d-1} d^{d-1}\hat{\Omega} }{(2 \pi)^{d+1}} \frac{1}{\omega^2 + c^2\ell^2 +2 c^2 k_F^2 (1-\cos\theta)} \frac{-1}{i \omega - v_F \ell + m} .
\ee
Wavefunction renormalization just depends on the linear term in $m$, so we can evaluate the simpler expression $\Sigma'(0)$.  The integral that results will show up in later diagrams as well, so it is convenient to define it here and calculate it once and for all:
\be \label{eq:FermionRenormIntegral}
{\cal I} \equiv 2\delta \Lambda \int \frac{d \ell k_F^{d-1}  d^{d-1} \hat{\Omega}}{(2\pi)^{d+1}} \frac{1}{\Lambda^2 + c^2 \ell^2 + 2 c^2 k_F^2 (1-\cos \theta)}\frac{1}{(i \Lambda -v_F \ell)^2}.
\ee
Performing the $\ell$ integral by contour integration and taking $d=3$, we find
\be
{\cal I} &=& -\frac{ \delta \Lambda}{2 \pi} \int_0^1 \frac{ 2k_F^2 d \sin^2 \frac{\theta}{2} }{(2 \pi)} 
\frac{c \sqrt{4c^2k_F^2 \sin^2 \frac{\theta}{2} + \Lambda^2}}{(4c^2k_F^2 \sin^2 \frac{\theta}{2} v_F + v_F \Lambda^2 + c \Lambda \sqrt{4c^2 k_F^2 \sin^2 \frac{\theta}{2} + \Lambda^2})^2} \stackrel{k_F \gg \Lambda}{=}- \frac{1 }{4 \pi^2 c v_F (c+v_F)} \frac{\delta \Lambda}{\Lambda}.
\ee
In terms of this integral, we simply have
\be
\Sigma'(0) &=& g^2 {\cal I} = - \frac{g^2 }{4 \pi^2 c v_F (c+v_F)} \frac{\delta \Lambda}{\Lambda}
\ee

Consequently, 
\be
a_g &=& \frac{1}{4\pi^2 c  v_F (c+v_F)}.
\ee

\subsection*{\begin{flushleft} (e) \end{flushleft}}
 
 The divergent part of the cubic renormalization is independent of the external momenta, so we can set these to zero.  The diagram then becomes
 \be
\delta g &=& g^3 \int \frac{d \omega d \ell d^{d-1} k_{\parallel} }{(2 \pi)^{d+1}} \frac{1}{\omega^2 + c^2 (k_\parallel^2 + \ell^2)} \frac{1}{(i \omega - v_F \ell )^2}
\nonumber
\ee
This is immediately seen to be $\delta g = g^3 {\cal I}$ from equation (\ref{eq:FermionRenormIntegral}), so we have
\be 
\delta g &=&  - \frac{g^3 }{4 \pi^2 c v_F (c+v_F)} \frac{\delta \Lambda}{\Lambda},
\ee
showing that $C_3=1$ in the theory with a single scalar $\phi$, as claimed in \S3.
 
\subsection*{\begin{flushleft} (f) \end{flushleft}}

The four-fermion vertex correction is just the usual BCS diagram, which we reproduce for completeness:
\be
\delta \lambda_\psi &=& \lambda_\psi^2 2 \delta \Lambda \int \frac{d\ell k_F^2 d^2 \hat{\Omega}}{(2\pi)^4}\frac{1}{(i \Lambda - v_F \ell)(-i \Lambda - v_F \ell)} = \frac{\lambda_\psi^2k_F^2}{4\pi^2 v_F} \frac{\delta \Lambda}{\Lambda}.
\ee
 There is an additional symmetry factor of $\frac{1}{2}$ for spin-less fermions.

\subsection*{\begin{flushleft} (g) and (i) \end{flushleft}}

As discussed in \S4, these diagrams produce $\log^2$ and $\log^3$ divergences.  A careful study of the meaning of these terms is beyond the scope of this article, and we will merely show below that they vanish at infinite $N$.

\subsection*{\begin{flushleft} (h) \end{flushleft}}

Finally, the diagram in  Fig. \ref{fig:alldiagrams}(h) gives no divergence due to the kinematic constraints of the fermi surface.  The explanation is similar to the reason that in the Fermi liquid theory with only forward scattering, there is no RG running.  To see this more clearly, here is the graph again, with momenta labeled:  

\begin{center}\includegraphics[width=0.18 \textwidth]{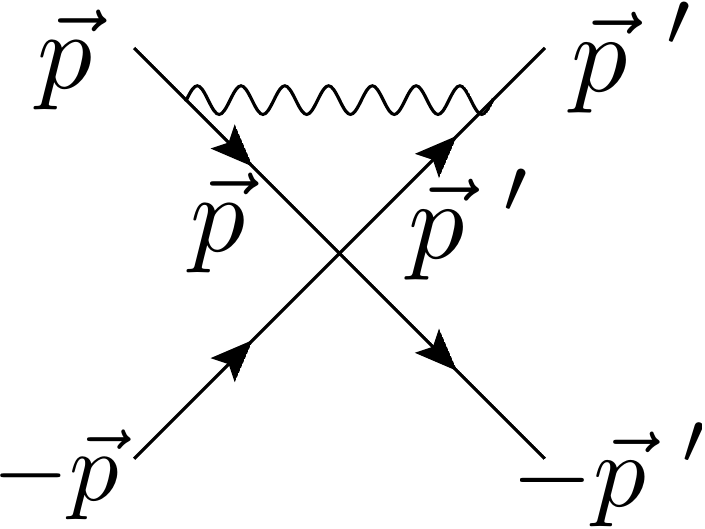}\end{center}

Because the insertion of the four-fermion interaction connects antipodal points, the internal fermion momenta must be $\vec{p}$ and $\vec{p}'$, as drawn.  However, the interaction being renormalized is also between antipodal points, so the external momenta at the top of the diagram must also be $\vec{p}$ and $\vec{p}'$ as drawn.  Therefore, the angular integral is constrained to a set of measure zero, and there is no UV divergence.    A similar, though slightly longer, argument applies to a diagram of the form of Fig. \ref{fig:alldiagrams}(i) rotated 90 degrees.

\subsection{Large N Generalization}
\label{app:matrix}

We now turn to calculating the $N$-dependent prefactors of the significant diagrams in Fig. \ref{fig:alldiagrams} in the $SU(N)$ generalization.  

\subsection*{\begin{flushleft} (a) \end{flushleft}}

Each cubic vertex brings down a factor of $\frac{g}{\sqrt{N}} \phi_A \bar{\psi}^i T^{Aj}_i \psi_j$.  Contracting the fermions to make a loop, one is left with $\frac{g^2}{N} \phi_A \phi_B {\rm tr}(T^A T^B) = \frac{1}{N} g^2 {\rm tr}(\phi^2)$, so the $N$-dependent prefactor is simply  $N^{-1}$.

\subsection*{\begin{flushleft} (c) (Matrix Valued Wilson-Fisher Fixed Point) \end{flushleft}}

The one-loop analysis of the Wilson-Fisher fixed point using the $\epsilon = 3 - d$ expansion has been a textbook subject for decades.  The story  changes  a little when we use the matrix valued $\phi_{ab}$ field; the  difference is that we need to study both tr$[\phi^2]^2$ and tr$[\phi^4]$ type terms:
\be
\CL_\phi \supset \frac{\lambda_\phi^{(1)}}{8 N} \tr (\phi^4) + \frac{\lambda_\phi^{(2)}}{8 N^2} (\tr (\phi^2))^2.
\ee 
  
Computing $SU(N)$ traces, we find that
\be
\beta_{\lambda^{(1)}} &=& \left[  \left( 1 - \frac{9}{N^2} \right) ( \lambda_\phi^{(1)})^2 + \frac{6}{N^2} \lambda_\phi^{(1)} \lambda_\phi^{(2)} \right] \beta_0 , \nn\\
\beta_{\lambda^{(2)}} &=& \left[ \left( \frac{3}{2} + \frac{9}{2N^2} \right) (\lambda_\phi^{(1)})^2 + 2 \left( 1-\frac{3}{2N^2} \right) \lambda_\phi^{(1)} \lambda_\phi^{(2)} + \frac{1}{2} \left( 1 + \frac{7}{N^2} \right) (\lambda_\phi^{(2)})^2 \right] \beta_0, \nn
\ee
where $\beta_0 = \frac{3}{16 \pi^2}$ is defined to be $\frac{\beta_\lambda}{\lambda^2}$ where $\beta_\lambda$ is the $\beta$ function of $\lambda$ in $\frac{\lambda}{4!} \phi^4$ theory.

This theory, with both couplings ${\lambda^{(1)}}$ and ${\lambda^{(2)}}$, does not have an IR-stable fixed point.  To see this, substitute the fixed point value $\lambda_\phi^{(1)} = \epsilon$ into $\beta_{\lambda^{(2)}}$ at leading order in large $N$ to give
\be
\beta_{\lambda^{(2)}} &=& \left[ \frac{3}{2} \epsilon^2 +  \epsilon \lambda_\phi^{(2)} + \frac{1}{2} (\lambda_\phi^{(2)})^2 \right] \beta_0
\ee
This function of $\lambda_\phi^{(2)}$ is always positive for all values of $\lambda_\phi^{(2)}$ and $\epsilon > 0$, which means that $\lambda_\phi^{(2)}$ will not have any fixed points, and will uncontrollably run negative.   This can be avoided by setting $\lambda_\phi^{(1)}= 0$, which is technically natural because when this coupling vanishes the bosonic sector has an enhanced $SO(N^2)$ symmetry.  In this case $\lambda_\phi^{(2)}$ has a fixed point for $\lambda_\phi^{(2)} = 2 \epsilon / \beta_0$.

\subsection*{\begin{flushleft} (d) \end{flushleft}}

For this diagram, the following identity for $SU(N)$ matrices $T^A$ (normalized so that ${\rm tr}(T^A T^B) = N \delta^{AB}$) proves useful:
\be
\frac{1}{N} T^{Ai}_j T^{Am}_n = \delta^i_n \delta^m_j - \frac{1}{N} \delta^i_j \delta^m_n.
\label{eq:genID}
\ee
Note that with our normalization of the $T^A$'s, the boson propagator is proportional to $1/N$. Contracting the fermions and using the above identity, we see that this diagram produces a factor of
\be
\frac{g}{\sqrt{N}}\frac{g}{\sqrt{N}} \bar{\psi}^i \psi_n ( N \delta^i_n - \frac{1}{N} \delta^i_n) = g^2 \bar{\psi}^i \psi_i \left( 1 - \frac{1}{N^2} \right),
\ee
so the $N$-dependence of the prefactor is simply $(1-N^{-2})$.  Consequently, the $SU(N)$ generalization of $a_g$ is
\be
a_g \rightarrow a_g^{(N)} = \left(1-\frac{1}{N^2}\right) a_g.
\ee

\subsection*{\begin{flushleft} (e) \end{flushleft}}

For the cubic vertex, we need the following identity for $SU(N)$ matrices:
\be
\sum_A T^A T^B T^A = - T^B.
\ee
The boson propagator introduces a factor of $1/N$ and each vertex introduces a factor of $\frac{g}{\sqrt{N}}$.  Therefore, $\frac{\delta g}{g}$ is proportional to $-\frac{1}{N^2}$.  Since $C_3 a_g$ was defined to be proportional to $-\frac{\delta g}{g^3}$, we therefore find
\be
C_3 = \frac{1}{N^2 -1}.
\ee

\subsection*{\begin{flushleft} (f) \end{flushleft}}

The BCS diagram renormalizing $\lambda_\psi$ does not involve any factors of the matrices $T^A$, and therefore all factors of $N$ are easily seen to come from the couplings themselves: $\frac{\delta \lambda_\psi}{\lambda_\psi} \propto \frac{\lambda_\psi^2 N^{-2}}{\lambda_\psi N^{-1}} = \lambda_\psi N^{-1}$. So in this case, the prefactor defined as $\beta$ in \S5 is just
\be
\beta = \frac{1}{N}
\ee

\subsection*{\begin{flushleft} (g) and (i) \end{flushleft}}

In both of these diagrams, boson exchanges just produce factors of $\frac{1}{N} T^{Ai}_j T^{Am}_n$, so the identity (\ref{eq:genID}) is all we need.  In fact, it is sufficient to note that there are no closed fermion loops in these diagrams, so there are no extra $N$-enhancements from tracing the identity matrix.  Therefore, the leading $N$-dependence is determined by the factors of $1/N$ from the couplings in the vertex insertions produce $\frac{\delta \lambda_\psi}{\lambda_\psi} \sim \frac{1}{N} g^2 \lambda_\psi$ and $\frac{1}{N} g^4$ for diagrams (g) and (i) respectively.  This is sufficient to suppress these diagrams at infinite $N$, as claimed.

\end{widetext}

\bibliography{qcmetal}

\begin{thebibliography}{37}%
\makeatletter
\providecommand \@ifxundefined [1]{%
 \@ifx{#1\undefined}
}%
\providecommand \@ifnum [1]{%
 \ifnum #1\expandafter \@firstoftwo
 \else \expandafter \@secondoftwo
 \fi
}%
\providecommand \@ifx [1]{%
 \ifx #1\expandafter \@firstoftwo
 \else \expandafter \@secondoftwo
 \fi
}%
\providecommand \natexlab [1]{#1}%
\providecommand \enquote  [1]{``#1''}%
\providecommand \bibnamefont  [1]{#1}%
\providecommand \bibfnamefont [1]{#1}%
\providecommand \citenamefont [1]{#1}%
\providecommand \href@noop [0]{\@secondoftwo}%
\providecommand \href [0]{\begingroup \@sanitize@url \@href}%
\providecommand \@href[1]{\@@startlink{#1}\@@href}%
\providecommand \@@href[1]{\endgroup#1\@@endlink}%
\providecommand \@sanitize@url [0]{\catcode `\\12\catcode `\$12\catcode
  `\&12\catcode `\#12\catcode `\^12\catcode `\_12\catcode `\%12\relax}%
\providecommand \@@startlink[1]{}%
\providecommand \@@endlink[0]{}%
\providecommand \url  [0]{\begingroup\@sanitize@url \@url }%
\providecommand \@url [1]{\endgroup\@href {#1}{\urlprefix }}%
\providecommand \urlprefix  [0]{URL }%
\providecommand \Eprint [0]{\href }%
\@ifxundefined \urlstyle {%
  \providecommand \doi  [0]{\begingroup \@sanitize@url \@doi}%
  \providecommand \@doi [1]{\endgroup \@@startlink {\doibase
  #1}doi:\discretionary {}{}{}#1\@@endlink }%
}{%
  \providecommand \doi  [0]{doi:\discretionary{}{}{}\begingroup
  \urlstyle{rm}\Url }%
}%
\providecommand \doibase [0]{http://dx.doi.org/}%
\providecommand \Doi [0]{\begingroup \@sanitize@url \@Doi }%
\providecommand \@Doi  [1]{\endgroup\@@startlink{\doibase#1}\@@Doi}%
\providecommand \@@Doi [1]{#1\@@endlink}%
\providecommand \selectlanguage [0]{\@gobble}%
\providecommand \bibinfo  [0]{\@secondoftwo}%
\providecommand \bibfield  [0]{\@secondoftwo}%
\providecommand \translation [1]{[#1]}%
\providecommand \BibitemOpen [0]{}%
\providecommand \bibitemStop [0]{}%
\providecommand \bibitemNoStop [0]{.\EOS\space}%
\providecommand \EOS [0]{\spacefactor3000\relax}%
\providecommand \BibitemShut  [1]{\csname bibitem#1\endcsname}%
\bibitem [{\citenamefont {Hertz}(1976)}]{Hertz1976}%
  \BibitemOpen
  \bibfield  {author} {\bibinfo {author} {\bibfnamefont {J.~A.}\ \bibnamefont
  {Hertz}},\ }\Doi {10.1103/PhysRevB.14.1165} {\bibfield  {journal} {\bibinfo
  {journal} {Phys. Rev. B},\ }\textbf {\bibinfo {volume} {14}},\ \bibinfo
  {pages} {1165} (\bibinfo {year} {1976})}\BibitemShut {NoStop}%
\bibitem [{\citenamefont {Millis}(1993)}]{Millis1993}%
  \BibitemOpen
  \bibfield  {author} {\bibinfo {author} {\bibfnamefont {A.~J.}\ \bibnamefont
  {Millis}},\ }\Doi {10.1103/PhysRevB.48.7183} {\bibfield  {journal} {\bibinfo
  {journal} {Phys. Rev. B},\ }\textbf {\bibinfo {volume} {48}},\ \bibinfo
  {pages} {7183} (\bibinfo {year} {1993})}\BibitemShut {NoStop}%
\bibitem [{\citenamefont {Landau}(1957)}]{Landau1957}%
  \BibitemOpen
  \bibfield  {author} {\bibinfo {author} {\bibfnamefont {L.}~\bibnamefont
  {Landau}},\ }\href@noop {} {\bibfield  {journal} {\bibinfo  {journal} {Sov.
  Phys. JETP},\ }\textbf {\bibinfo {volume} {3}},\ \bibinfo {pages} {920}
  (\bibinfo {year} {1957})}\BibitemShut {NoStop}%
\bibitem [{\citenamefont {Shankar}(1991)}]{Shankar1991}%
  \BibitemOpen
  \bibfield  {author} {\bibinfo {author} {\bibfnamefont {R.}~\bibnamefont
  {Shankar}},\ }\href@noop {} {\bibfield  {journal} {\bibinfo  {journal}
  {Physica A: Statistical Mechanics and its Applications},\ }\textbf {\bibinfo
  {volume} {177}},\ \bibinfo {pages} {530} (\bibinfo {year}
  {1991})}\BibitemShut {NoStop}%
\bibitem [{\citenamefont {Polchinski}(1992)}]{Polchinski1992}%
  \BibitemOpen
  \bibfield  {author} {\bibinfo {author} {\bibfnamefont {J.}~\bibnamefont
  {Polchinski}},\ }\href@noop {} {\bibfield  {journal} {\bibinfo  {journal}
  {arXiv:hep-th 9210046, TASI 1992 Lectures}} (\bibinfo {year}
  {1992})}\BibitemShut {NoStop}%
\bibitem [{\citenamefont {Shankar}(1994)}]{Shankar1994}%
  \BibitemOpen
  \bibfield  {author} {\bibinfo {author} {\bibfnamefont {R.}~\bibnamefont
  {Shankar}},\ }\href@noop {} {\bibfield  {journal} {\bibinfo  {journal} {Rev.
  Mod. Phys.},\ }\textbf {\bibinfo {volume} {66}},\ \bibinfo {pages} {129}
  (\bibinfo {year} {1994})}\BibitemShut {NoStop}%
\bibitem [{\citenamefont {L\"ohneysen}\ \emph {et~al.}(2007)\citenamefont
  {L\"ohneysen}, \citenamefont {Rosch}, \citenamefont {Vojta},\ and\
  \citenamefont {W\"olfle}}]{Lohneysen2007}%
  \BibitemOpen
  \bibfield  {author} {\bibinfo {author} {\bibfnamefont {H.~v.}\ \bibnamefont
  {L\"ohneysen}}, \bibinfo {author} {\bibfnamefont {A.}~\bibnamefont {Rosch}},
  \bibinfo {author} {\bibfnamefont {M.}~\bibnamefont {Vojta}}, \ and\ \bibinfo
  {author} {\bibfnamefont {P.}~\bibnamefont {W\"olfle}},\ }\Doi
  {10.1103/RevModPhys.79.1015} {\bibfield  {journal} {\bibinfo  {journal} {Rev.
  Mod. Phys.},\ }\textbf {\bibinfo {volume} {79}},\ \bibinfo {pages} {1015}
  (\bibinfo {year} {2007})}\BibitemShut {NoStop}%
\bibitem [{\citenamefont {Stewart}(2001)}]{Stewart2001}%
  \BibitemOpen
  \bibfield  {author} {\bibinfo {author} {\bibfnamefont {G.~R.}\ \bibnamefont
  {Stewart}},\ }\Doi {10.1103/RevModPhys.73.797} {\bibfield  {journal}
  {\bibinfo  {journal} {Rev. Mod. Phys.},\ }\textbf {\bibinfo {volume} {73}},\
  \bibinfo {pages} {797} (\bibinfo {year} {2001})}\BibitemShut {NoStop}%
\bibitem [{\citenamefont {Holstein}\ \emph {et~al.}(1973)\citenamefont
  {Holstein}, \citenamefont {Norton},\ and\ \citenamefont {Pincus}}]{Holstein}%
  \BibitemOpen
  \bibfield  {author} {\bibinfo {author} {\bibfnamefont {T.}~\bibnamefont
  {Holstein}}, \bibinfo {author} {\bibfnamefont {R.}~\bibnamefont {Norton}}, \
  and\ \bibinfo {author} {\bibfnamefont {P.}~\bibnamefont {Pincus}},\
  }\href@noop {} {\bibfield  {journal} {\bibinfo  {journal} {Phys. Rev. B},\
  }\textbf {\bibinfo {volume} {8}},\ \bibinfo {pages} {2649} (\bibinfo {year}
  {1973})}\BibitemShut {NoStop}%
\bibitem [{\citenamefont {Varma}\ \emph {et~al.}(1989)\citenamefont {Varma},
  \citenamefont {Littlewood}, \citenamefont {Schmitt-Rink}, \citenamefont
  {Abrahams},\ and\ \citenamefont {Ruckenstein}}]{Varma}%
  \BibitemOpen
  \bibfield  {author} {\bibinfo {author} {\bibfnamefont {C.}~\bibnamefont
  {Varma}}, \bibinfo {author} {\bibfnamefont {P.}~\bibnamefont {Littlewood}},
  \bibinfo {author} {\bibfnamefont {S.}~\bibnamefont {Schmitt-Rink}}, \bibinfo
  {author} {\bibfnamefont {E.}~\bibnamefont {Abrahams}}, \ and\ \bibinfo
  {author} {\bibfnamefont {A.}~\bibnamefont {Ruckenstein}},\ }\href@noop {}
  {\bibfield  {journal} {\bibinfo  {journal} {Phys. Rev. Lett.},\ }\textbf
  {\bibinfo {volume} {63}},\ \bibinfo {pages} {1996} (\bibinfo {year}
  {1989})}\BibitemShut {NoStop}%
\bibitem [{\citenamefont {Altshuler}\ \emph {et~al.}(1994)\citenamefont
  {Altshuler}, \citenamefont {Ioffe},\ and\ \citenamefont
  {Millis}}]{Altshuler1994}%
  \BibitemOpen
  \bibfield  {author} {\bibinfo {author} {\bibfnamefont {B.~L.}\ \bibnamefont
  {Altshuler}}, \bibinfo {author} {\bibfnamefont {L.~B.}\ \bibnamefont
  {Ioffe}}, \ and\ \bibinfo {author} {\bibfnamefont {A.~J.}\ \bibnamefont
  {Millis}},\ }\Doi {10.1103/PhysRevB.50.14048} {\bibfield  {journal} {\bibinfo
   {journal} {Phys. Rev. B},\ }\textbf {\bibinfo {volume} {50}},\ \bibinfo
  {pages} {14048} (\bibinfo {year} {1994})}\BibitemShut {NoStop}%
\bibitem [{\citenamefont {Nayak}\ and\ \citenamefont
  {Wilczek}(1994){\natexlab{a}}}]{Nayak1994}%
  \BibitemOpen
  \bibfield  {author} {\bibinfo {author} {\bibfnamefont {C.}~\bibnamefont
  {Nayak}}\ and\ \bibinfo {author} {\bibfnamefont {F.}~\bibnamefont
  {Wilczek}},\ }\href@noop {} {\bibfield  {journal} {\bibinfo  {journal}
  {Nuclear Physics B},\ }\textbf {\bibinfo {volume} {417}},\ \bibinfo {pages}
  {359} (\bibinfo {year} {1994}{\natexlab{a}})}\BibitemShut {NoStop}%
\bibitem [{\citenamefont {Polchinski}(1994)}]{Polchinski1994}%
  \BibitemOpen
  \bibfield  {author} {\bibinfo {author} {\bibfnamefont {J.}~\bibnamefont
  {Polchinski}},\ }\href@noop {} {\bibfield  {journal} {\bibinfo  {journal}
  {Nuclear Physics B},\ }\textbf {\bibinfo {volume} {422}},\ \bibinfo {pages}
  {617} (\bibinfo {year} {1994})}\BibitemShut {NoStop}%
\bibitem [{\citenamefont {Chakravarty}\ \emph {et~al.}(1995)\citenamefont
  {Chakravarty}, \citenamefont {Norton},\ and\ \citenamefont
  {Syljuasen}}]{Chakravarty1995}%
  \BibitemOpen
  \bibfield  {author} {\bibinfo {author} {\bibfnamefont {S.}~\bibnamefont
  {Chakravarty}}, \bibinfo {author} {\bibfnamefont {R.}~\bibnamefont {Norton}},
  \ and\ \bibinfo {author} {\bibfnamefont {O.}~\bibnamefont {Syljuasen}},\
  }\href@noop {} {\bibfield  {journal} {\bibinfo  {journal} {Phys. Rev.
  Lett.},\ }\textbf {\bibinfo {volume} {74}},\ \bibinfo {pages} {1423}
  (\bibinfo {year} {1995})}\BibitemShut {NoStop}%
\bibitem [{\citenamefont {Oganesyan}\ \emph {et~al.}(2001)\citenamefont
  {Oganesyan}, \citenamefont {Kivelson},\ and\ \citenamefont
  {Fradkin}}]{Oganesyan2001}%
  \BibitemOpen
  \bibfield  {author} {\bibinfo {author} {\bibfnamefont {V.}~\bibnamefont
  {Oganesyan}}, \bibinfo {author} {\bibfnamefont {S.~A.}\ \bibnamefont
  {Kivelson}}, \ and\ \bibinfo {author} {\bibfnamefont {E.}~\bibnamefont
  {Fradkin}},\ }\Doi {10.1103/PhysRevB.64.195109} {\bibfield  {journal}
  {\bibinfo  {journal} {Phys. Rev. B},\ }\textbf {\bibinfo {volume} {64}},\
  \bibinfo {pages} {195109} (\bibinfo {year} {2001})}\BibitemShut {NoStop}%
\bibitem [{\citenamefont {Varma}\ \emph {et~al.}(2002)\citenamefont {Varma},
  \citenamefont {Nussinov},\ and\ \citenamefont {van Saarloos}}]{Varma2002}%
  \BibitemOpen
  \bibfield  {author} {\bibinfo {author} {\bibfnamefont {C.}~\bibnamefont
  {Varma}}, \bibinfo {author} {\bibfnamefont {Z.}~\bibnamefont {Nussinov}}, \
  and\ \bibinfo {author} {\bibfnamefont {W.}~\bibnamefont {van Saarloos}},\
  }\href@noop {} {\bibfield  {journal} {\bibinfo  {journal} {Physics Reports},\
  }\textbf {\bibinfo {volume} {361}},\ \bibinfo {pages} {267} (\bibinfo {year}
  {2002})}\BibitemShut {NoStop}%
\bibitem [{\citenamefont {Lawler}\ and\ \citenamefont
  {Fradkin}(2007)}]{FradkinExtra}%
  \BibitemOpen
  \bibfield  {author} {\bibinfo {author} {\bibfnamefont {M.}~\bibnamefont
  {Lawler}}\ and\ \bibinfo {author} {\bibfnamefont {E.}~\bibnamefont
  {Fradkin}},\ }\Doi {10.1103/PhysRevB.75.033304} {\bibfield  {journal}
  {\bibinfo  {journal} {Phys. Rev. B},\ }\textbf {\bibinfo {volume} {75}},\
  \bibinfo {pages} {033304} (\bibinfo {year} {2007})}\BibitemShut {NoStop}%
\bibitem [{\citenamefont {Senthil}\ and\ \citenamefont
  {Shankar}(2009)}]{Senthil2009}%
  \BibitemOpen
  \bibfield  {author} {\bibinfo {author} {\bibfnamefont {T.}~\bibnamefont
  {Senthil}}\ and\ \bibinfo {author} {\bibfnamefont {R.}~\bibnamefont
  {Shankar}},\ }\Doi {10.1103/PhysRevLett.102.046406} {\bibfield  {journal}
  {\bibinfo  {journal} {Phys. Rev. Lett.},\ }\textbf {\bibinfo {volume}
  {102}},\ \bibinfo {pages} {046406} (\bibinfo {year} {2009})}\BibitemShut
  {NoStop}%
\bibitem [{\citenamefont {Lee}(2009)}]{Lee2009}%
  \BibitemOpen
  \bibfield  {author} {\bibinfo {author} {\bibfnamefont {S.-S.}\ \bibnamefont
  {Lee}},\ }\Doi {10.1103/PhysRevB.80.165102} {\bibfield  {journal} {\bibinfo
  {journal} {Phys. Rev. B},\ }\textbf {\bibinfo {volume} {80}},\ \bibinfo
  {pages} {165102} (\bibinfo {year} {2009})}\BibitemShut {NoStop}%
\bibitem [{\citenamefont {Metlitski}\ and\ \citenamefont
  {Sachdev}(2010)}]{Metlitski2010}%
  \BibitemOpen
  \bibfield  {author} {\bibinfo {author} {\bibfnamefont {M.~A.}\ \bibnamefont
  {Metlitski}}\ and\ \bibinfo {author} {\bibfnamefont {S.}~\bibnamefont
  {Sachdev}},\ }\Doi {10.1103/PhysRevB.82.075127} {\bibfield  {journal}
  {\bibinfo  {journal} {Phys. Rev. B},\ }\textbf {\bibinfo {volume} {82}},\
  \bibinfo {pages} {075127} (\bibinfo {year} {2010})}\BibitemShut {NoStop}%
\bibitem [{\citenamefont {Mross}\ \emph {et~al.}(2010)\citenamefont {Mross},
  \citenamefont {McGreevy}, \citenamefont {Liu},\ and\ \citenamefont
  {Senthil}}]{Mross2010}%
  \BibitemOpen
  \bibfield  {author} {\bibinfo {author} {\bibfnamefont {D.~F.}\ \bibnamefont
  {Mross}}, \bibinfo {author} {\bibfnamefont {J.}~\bibnamefont {McGreevy}},
  \bibinfo {author} {\bibfnamefont {H.}~\bibnamefont {Liu}}, \ and\ \bibinfo
  {author} {\bibfnamefont {T.}~\bibnamefont {Senthil}},\ }\Doi
  {10.1103/PhysRevB.82.045121} {\bibfield  {journal} {\bibinfo  {journal}
  {Phys. Rev. B},\ }\textbf {\bibinfo {volume} {82}},\ \bibinfo {pages}
  {045121} (\bibinfo {year} {2010})}\BibitemShut {NoStop}%
\bibitem [{\citenamefont {{Dalidovich}}\ and\ \citenamefont
  {{Lee}}(2013)}]{Lee2013}%
  \BibitemOpen
  \bibfield  {author} {\bibinfo {author} {\bibfnamefont {D.}~\bibnamefont
  {{Dalidovich}}}\ and\ \bibinfo {author} {\bibfnamefont {S.-S.}\ \bibnamefont
  {{Lee}}},\ }\href@noop {} {\bibfield  {journal} {\bibinfo  {journal} {ArXiv
  e-prints}} (\bibinfo {year} {2013})},\ \Eprint
  {http://arxiv.org/abs/1307.3170} {arXiv:1307.3170 [cond-mat.str-el]}
  \BibitemShut {NoStop}%
\bibitem [{\citenamefont {Abanov}\ and\ \citenamefont
  {Chubukov}(2004)}]{Abanov2004}%
  \BibitemOpen
  \bibfield  {author} {\bibinfo {author} {\bibfnamefont {A.}~\bibnamefont
  {Abanov}}\ and\ \bibinfo {author} {\bibfnamefont {A.}~\bibnamefont
  {Chubukov}},\ }\Doi {10.1103/PhysRevLett.93.255702} {\bibfield  {journal}
  {\bibinfo  {journal} {Phys. Rev. Lett.},\ }\textbf {\bibinfo {volume} {93}},\
  \bibinfo {pages} {255702} (\bibinfo {year} {2004})}\BibitemShut {NoStop}%
\bibitem [{\citenamefont {Belitz}\ \emph {et~al.}(2005)\citenamefont {Belitz},
  \citenamefont {Kirkpatrick},\ and\ \citenamefont {Vojta}}]{Belitz2005}%
  \BibitemOpen
  \bibfield  {author} {\bibinfo {author} {\bibfnamefont {D.}~\bibnamefont
  {Belitz}}, \bibinfo {author} {\bibfnamefont {T.~R.}\ \bibnamefont
  {Kirkpatrick}}, \ and\ \bibinfo {author} {\bibfnamefont {T.}~\bibnamefont
  {Vojta}},\ }\Doi {10.1103/RevModPhys.77.579} {\bibfield  {journal} {\bibinfo
  {journal} {Rev. Mod. Phys.},\ }\textbf {\bibinfo {volume} {77}},\ \bibinfo
  {pages} {579} (\bibinfo {year} {2005})}\BibitemShut {NoStop}%
\bibitem [{\citenamefont {Sachdev}(2011)}]{Sachdev}%
  \BibitemOpen
  \bibfield  {author} {\bibinfo {author} {\bibfnamefont {S.}~\bibnamefont
  {Sachdev}},\ }\href@noop {} {\emph {\bibinfo {title} {Quantum Phase
  Transitions}}}\ (\bibinfo  {publisher} {Cambridge University Press},\
  \bibinfo {address} {Cambridge},\ \bibinfo {year} {2011})\BibitemShut
  {NoStop}%
\bibitem [{\citenamefont {Fradkin}\ \emph {et~al.}(2010)\citenamefont
  {Fradkin}, \citenamefont {Kivelson}, \citenamefont {Lawler}, \citenamefont
  {Eisenstein},\ and\ \citenamefont {Mackenzie}}]{Fradkin2010}%
  \BibitemOpen
  \bibfield  {author} {\bibinfo {author} {\bibfnamefont {E.}~\bibnamefont
  {Fradkin}}, \bibinfo {author} {\bibfnamefont {S.~A.}\ \bibnamefont
  {Kivelson}}, \bibinfo {author} {\bibfnamefont {M.~J.}\ \bibnamefont
  {Lawler}}, \bibinfo {author} {\bibfnamefont {J.~P.}\ \bibnamefont
  {Eisenstein}}, \ and\ \bibinfo {author} {\bibfnamefont {A.~P.}\ \bibnamefont
  {Mackenzie}},\ }\Doi {10.1146/annurev-conmatphys-070909-103925} {\bibfield
  {journal} {\bibinfo  {journal} {Annual Review of Condensed Matter Physics},\
  }\textbf {\bibinfo {volume} {1}},\ \bibinfo {pages} {153} (\bibinfo {year}
  {2010})}\BibitemShut {NoStop}%
\bibitem [{\citenamefont {Lee}(2008)}]{Lee2008}%
  \BibitemOpen
  \bibfield  {author} {\bibinfo {author} {\bibfnamefont {S.-S.}\ \bibnamefont
  {Lee}},\ }\Doi {10.1103/PhysRevB.78.085129} {\bibfield  {journal} {\bibinfo
  {journal} {Phys. Rev. B},\ }\textbf {\bibinfo {volume} {78}},\ \bibinfo
  {pages} {085129} (\bibinfo {year} {2008})}\BibitemShut {NoStop}%
\bibitem [{\citenamefont {{Phillips}}\ \emph {et~al.}(2013)\citenamefont
  {{Phillips}}, \citenamefont {{Langley}},\ and\ \citenamefont
  {{Hutasoit}}}]{Phillips2013}%
  \BibitemOpen
  \bibfield  {author} {\bibinfo {author} {\bibfnamefont {P.~W.}\ \bibnamefont
  {{Phillips}}}, \bibinfo {author} {\bibfnamefont {B.~W.}\ \bibnamefont
  {{Langley}}}, \ and\ \bibinfo {author} {\bibfnamefont {J.~A.}\ \bibnamefont
  {{Hutasoit}}},\ }\href@noop {} {\bibfield  {journal} {\bibinfo  {journal}
  {ArXiv e-prints}} (\bibinfo {year} {2013})},\ \Eprint
  {http://arxiv.org/abs/1305.0006} {arXiv:1305.0006 [cond-mat.str-el]}
  \BibitemShut {NoStop}%
\bibitem [{\citenamefont {Nayak}\ and\ \citenamefont
  {Wilczek}(1994){\natexlab{b}}}]{Nayak1994a}%
  \BibitemOpen
  \bibfield  {author} {\bibinfo {author} {\bibfnamefont {C.}~\bibnamefont
  {Nayak}}\ and\ \bibinfo {author} {\bibfnamefont {F.}~\bibnamefont
  {Wilczek}},\ }\href@noop {} {\bibfield  {journal} {\bibinfo  {journal}
  {Nuclear Physics B},\ }\textbf {\bibinfo {volume} {430}},\ \bibinfo {pages}
  {534} (\bibinfo {year} {1994}{\natexlab{b}})}\BibitemShut {NoStop}%
\bibitem [{\citenamefont {Son}(1999)}]{Son1999}%
  \BibitemOpen
  \bibfield  {author} {\bibinfo {author} {\bibfnamefont {D.~T.}\ \bibnamefont
  {Son}},\ }\Doi {10.1103/PhysRevD.59.094019} {\bibfield  {journal} {\bibinfo
  {journal} {Phys. Rev. D},\ }\textbf {\bibinfo {volume} {59}},\ \bibinfo
  {pages} {094019} (\bibinfo {year} {1999})}\BibitemShut {NoStop}%
\bibitem [{\citenamefont {Sch\"afer}\ and\ \citenamefont
  {Wilczek}(1999)}]{Schafer1999}%
  \BibitemOpen
  \bibfield  {author} {\bibinfo {author} {\bibfnamefont {T.}~\bibnamefont
  {Sch\"afer}}\ and\ \bibinfo {author} {\bibfnamefont {F.}~\bibnamefont
  {Wilczek}},\ }\Doi {10.1103/PhysRevD.60.114033} {\bibfield  {journal}
  {\bibinfo  {journal} {Phys. Rev. D},\ }\textbf {\bibinfo {volume} {60}},\
  \bibinfo {pages} {114033} (\bibinfo {year} {1999})}\BibitemShut {NoStop}%
\bibitem [{\citenamefont {Sch{\"a}fer}(2001)}]{Schafer2001}%
  \BibitemOpen
  \bibfield  {author} {\bibinfo {author} {\bibfnamefont {T.}~\bibnamefont
  {Sch{\"a}fer}},\ }\href@noop {} {\bibfield  {journal} {\bibinfo  {journal}
  {International Journal of Modern Physics B},\ }\textbf {\bibinfo {volume}
  {15}},\ \bibinfo {pages} {1474} (\bibinfo {year} {2001})}\BibitemShut
  {NoStop}%
\bibitem [{\citenamefont {Wang}\ and\ \citenamefont
  {Chubukov}(2013)}]{Wang2013}%
  \BibitemOpen
  \bibfield  {author} {\bibinfo {author} {\bibfnamefont {Y.}~\bibnamefont
  {Wang}}\ and\ \bibinfo {author} {\bibfnamefont {A.~V.}\ \bibnamefont
  {Chubukov}},\ }\Doi {10.1103/PhysRevLett.110.127001} {\bibfield  {journal}
  {\bibinfo  {journal} {Phys. Rev. Lett.},\ }\textbf {\bibinfo {volume}
  {110}},\ \bibinfo {pages} {127001} (\bibinfo {year} {2013})}\BibitemShut
  {NoStop}%
\bibitem [{\citenamefont {{Mahajan}}\ \emph {et~al.}(2013)\citenamefont
  {{Mahajan}}, \citenamefont {{Ramirez}}, \citenamefont {{Kachru}},\ and\
  \citenamefont {{Raghu}}}]{Mahajan2013}%
  \BibitemOpen
  \bibfield  {author} {\bibinfo {author} {\bibfnamefont {R.}~\bibnamefont
  {{Mahajan}}}, \bibinfo {author} {\bibfnamefont {D.~M.}\ \bibnamefont
  {{Ramirez}}}, \bibinfo {author} {\bibfnamefont {S.}~\bibnamefont {{Kachru}}},
  \ and\ \bibinfo {author} {\bibfnamefont {S.}~\bibnamefont {{Raghu}}},\
  }\href@noop {} {\bibfield  {journal} {\bibinfo  {journal} {ArXiv e-prints}}
  (\bibinfo {year} {2013})},\ \Eprint {http://arxiv.org/abs/1303.1587}
  {arXiv:1303.1587 [cond-mat.str-el]} \BibitemShut {NoStop}%
\bibitem [{Note1()}]{Note1}%
  \BibitemOpen
  \bibinfo {note} {A different class of diagrams contributing to the overlap
  region between forward and antipodal scattering is not suppressed at large
  $N$, and was considered in [\protect \rev@citealpnum {SonShuster}]. The
  interesting effects of these diagrams set in at a scale that is exponentially
  small at small $\epsilon $, and we will not discuss them
  further.}\BibitemShut {Stop}%
\bibitem [{\citenamefont {Chung}\ \emph {et~al.}(2013)\citenamefont {Chung},
  \citenamefont {Mandal}, \citenamefont {Raghu},\ and\ \citenamefont
  {Chakravarty}}]{Chung2013}%
  \BibitemOpen
  \bibfield  {author} {\bibinfo {author} {\bibfnamefont {S.-B.}\ \bibnamefont
  {Chung}}, \bibinfo {author} {\bibfnamefont {I.}~\bibnamefont {Mandal}},
  \bibinfo {author} {\bibfnamefont {S.}~\bibnamefont {Raghu}}, \ and\ \bibinfo
  {author} {\bibfnamefont {S.}~\bibnamefont {Chakravarty}},\ }\href@noop {}
  {\bibfield  {journal} {\bibinfo  {journal} {arXiv:1305.3938}} (\bibinfo
  {year} {2013})}\BibitemShut {NoStop}%
\bibitem [{\citenamefont {Shuster}\ and\ \citenamefont
  {Son}(2000)}]{SonShuster}%
  \BibitemOpen
  \bibfield  {author} {\bibinfo {author} {\bibfnamefont {E.}~\bibnamefont
  {Shuster}}\ and\ \bibinfo {author} {\bibfnamefont {D.}~\bibnamefont {Son}},\
  }\Doi {10.1016/S0550-3213(99)00615-X} {\bibfield  {journal} {\bibinfo
  {journal} {Nucl.Phys.},\ }\textbf {\bibinfo {volume} {B573}},\ \bibinfo
  {pages} {434} (\bibinfo {year} {2000})},\ \Eprint
  {http://arxiv.org/abs/hep-ph/9905448} {arXiv:hep-ph/9905448 [hep-ph]}
  \BibitemShut {NoStop}%
\end{thebibliography}%

\end{document}